\documentclass{aa}
\usepackage{txfonts}
\usepackage{graphicx}
\usepackage{natbib}
\usepackage{multirow}
\bibpunct{(}{)}{;}{a}{}{,} % to follow the A&A style
\newcommand{\micron}{\mbox{$\mu$m}}

\newcommand{\subscr}[1]{\ensuremath{_{\textrm{\scriptsize{#1}}}}}
\newcommand{\trot}{\mbox{$T_{\rm rot}$}}
\newcommand{\tk}{\mbox{$T_{K}$}}

\begin{document}

	\title{DIGIT:  \textit{Herschel}\thanks{{\it Herschel} is an ESA space
observatory with science instruments provided by European-led Principal
Investigator consortia and with important participation from NASA.} and
\textit{Spitzer} spectro-imaging of \\ SMM3 and SMM4 in Serpens}
\author{O.~Dionatos\inst{1,2,3} \and J.~K.~J{\o}rgensen \inst{2,1} \and
J.~D.~Green \inst{4} \and G.~J.~Herczeg \inst{5} \and N.~J.~Evans II \inst{4}
\and L.~E.~Kristensen \inst{6} \and J.~E.~Lindberg \inst{1,2} \and E.~F.~van
Dishoeck \inst{6,7} }

\institute{ Centre for Star and Planet Formation, Natural History Museum of
Denmark, University of Copenhagen,  {\O}ster Voldgade 5 -- 7, DK-1350 Copenhagen K.
Denmark\\ \email{odysseas@nbi.ku.dk} \\ \and Niels Bohr Institute, University of
Copenhagen. Juliane Maries Vej 30, DK-2100 Copenhagen {\O}. Denmark\\ \and
University of Vienna, Department of Astronomy,  T{\"u}rkenschanzstrasse 17,
A-1180, Vienna, Austria\\ \and University of Texas at Austin, Department of
Astronomy, 2515 Speedway, Stop C1400, Austin, TX 78712-1205, USA\\ \and Kavli
Institute for Astronomy and Astrophysics, Peking University, Beijing, 100871, PR
China\\ \and Leiden Observatory, Leiden University, P.O. Box 9513, NL-2300 RA
Leiden, The Netherlands\\ \and Max Planck Institut f{\"u}r Extraterrestrische
Physik, Giessenbachstrasse, D-85748 Garching, Germany\\ }

%\titilerunning
\abstract
% context
{Mid- and far-infrared observations of the environment around embedded
protostars reveal a plethora of high excitation molecular and atomic emission
lines. Different mechanisms for the origin of these lines have been proposed,
including shocks induced by protostellar jets, and radiation or heating by the
embedded protostar of its immediate surroundings.}
% aims
{Through the study of the most important molecular and atomic coolants, we aim
at constraining the physical conditions around two protostars in the Serpens
molecular cloud core and measuring the CO/H$_2$ ratio in warm gas.}
% methods
{We employ spectro-imaging observations from Spitzer/IRS and Herschel/PACS
providing an almost complete wavelength coverage between 5 and 200 $\micron$ to
observe the environment around the embedded protostars SMM3 and SMM4 in Serpens.
Within this range, emission from all major molecular (H$_2$, CO, H$_2$O and OH)
and many atomic ([OI], [CII], [FeII], [SiII] and [SI]) coolants of excited gas
are detected. } %Emission line maps reveal the morphology of the observed emission
%and indicate associations between the different species. The excitation
%conditions for molecular species are assessed through rotational diagrams.
%Emission lines from major coolants are compared to the results of steady-state
%C- and J-type shock models.}
%results
{Line emission tends to peak at distances of $\sim$10--20$\arcsec$  from the
protostellar sources, with all but [CII] peaking at the positions of outflow
shocks seen in near-IR and sub-millimeter interferometric observations. 
%The
%[CII] emission pattern suggests that it is most likely excited from energetic UV
%radiation originating from the nearby flat-spectrum source SMM6. 
Excitation
analysis indicates that H$_2$ and CO originate from gas at two distinct
rotational temperatures of $\sim$300 K and 1000 K, while the excitation
temperature for H$_2$O and OH is $\sim$100--200 K. The morphological and
physical association between CO and H$_2$ suggests a common excitation mechanism
which allows direct comparisons between the two molecules. The CO/H$_2$
abundance ratio varies from $\sim 10^{-5}$ in the warmer gas up to $\sim
10^{-4}$ in the hotter regions. %Shock models indicate that C-shocks can account
%for the observed line intensities if a beam filling factor and a temperature
%stratification in the shock front are considered. C-type shocks can best explain
%the emission from H$_2$O. 
The existence of J-shocks is suggested by the strong
atomic/ionic (except for [CII]) emission as well as a number of line ratio
diagnostics.  Dissociative shocks can account for the CO and H$_2$ emission in a
single excitation temperature structure.}
%conclusions
{The bulk of cooling from molecular and atomic lines is associated with gas
excited in outflow shocks. The strong association between H$_2$ and CO enable to
constrain their abundance ratio in warm gas, which is found to vary between
10$^{-4}$ and 10$^{-5}$. Both C- and J-type shocks can account for the observed
molecular emission, however J-shocks are strongly advocated by the atomic
emission and provide simpler and more homogeneous solutions for CO, H$_2$. The
variations in the CO/H$_2$ abundance ratio for gas at different temperatures can
be interpreted by their reformation rates in dissociative J-type shocks, or the
influence of both C and J shocks}


\maketitle

\section {Introduction} \label{sec:1}

Protostellar outflows moving at supersonic velocities interact with the
interstellar medium through shocks. Such interactions result in heating,
compressing and setting the gas into motion.  On large, outflow scales, varying
physical conditions may occur due to the intrinsic physical properties of the
underlying primary protostellar jet and the way it propagates
\citep[e.g.][]{Arce:07a}. On scales of individual shocks, the physics depends
mainly on the energy transfer and the possible presence of magnetic fields
\citep[e.g.][]{Hollenbach:89a}.  The combined effect of these mechanisms produce
physical and chemical gradients along the outflows on all spatial scales.
Shock-excited gas cools mainly through atomic and molecular line radiation,
e.g., near- and mid-infrared H$_2$ transitions (corresponding to temperatures of
a few thousand K) and far-infrared lines of CO and H$_2$O (temperatures from a
few hundred up to $\sim$~1000~K). Swept-up ambient material with temperatures of
a few tens of K is commonly traced by low energy CO transitions, falling in the
(sub)millimeter regime \citep{Arce:07a}. A complete census connecting these two
regimes is required for understanding the physical mechanisms relating the
shocks to the large-scale outflows and the energy transfer from the vicinity of
the protostar to the surrounding medium, including the cooling taking place at
the intermediate energies. 

The ISO satellite \citep{Kessler:96a} made important contributions bridging the
two energy regimes \citep[e.g.][]{van-Dishoeck:04a}. Observations with the SWS,
LWS and CVF instruments revealed many molecular (H$_2$, CO, H$_2$O, OH) and
atomic ([OI], [CII]) lines, which have helped to understand the shock cooling in
protostellar outflows \citep[][]{Giannini:01a}. The angular resolution of ISO,
however, was only sufficient to study the cumulative effect of different
processes within its beam of $\sim 80 \arcsec$ (LWS) and understanding the
outflow physics therefore had to rely on in-depth modeling
\citep[e.g.][]{Nisini:00a, Giannini:06a}.

Observations of protostellar outflows with the Spitzer Space Telescope
\citep{Werner:04a} and more recently the Herschel Space Observatory
\citep{Pilbratt:10a} have provided us a more detailed view of the physical
processes along outflows, through spectral images with angular resolutions
ranging between 3.5$\arcsec$ and 9.4$\arcsec$ (for the IRS and PACS instruments,
respectively).  Spectro-imaging is a powerful tool for the study of extended
emission structures, permitting us to retrieve simultaneously the spatial
distribution and spectral information of an excited region. This allows to
disentangle the contribution of UV-heated outflow cavities from emission due to
shocks along the outflow propagation axis \citep[e.g.][]{Visser:12a,
Herczeg:12a, van-Kempen:10b}. The spatial distribution of isolated spectral
features provides morphological evidence for the origin of the excited gas,
whereas the study of the spectral information provides information on the
underlying physical conditions \citep[e.g.][]{Kristensen:10a}. In the mid-IR,
``synthetic" (i.e. reconstructed from slit-scan observations) spectral maps have
been obtained with the Spitzer/IRS spectrograph for a number of individual
protostars \citep{Neufeld:06a, Neufeld:09a, Dionatos:10b, Nisini:10a} and
star-forming regions \citep{Maret:09a}. PACS onboard Herschel comes with
built-in spectro-imaging capabilities at angular resolutions comparable to
Spitzer/IRS. We here employ spectral maps from both instruments to study the
excitation of the medium around the embedded protostars in the Serpens star
forming region.

\begin{figure} \centering
\resizebox{\hsize}{!}{\includegraphics{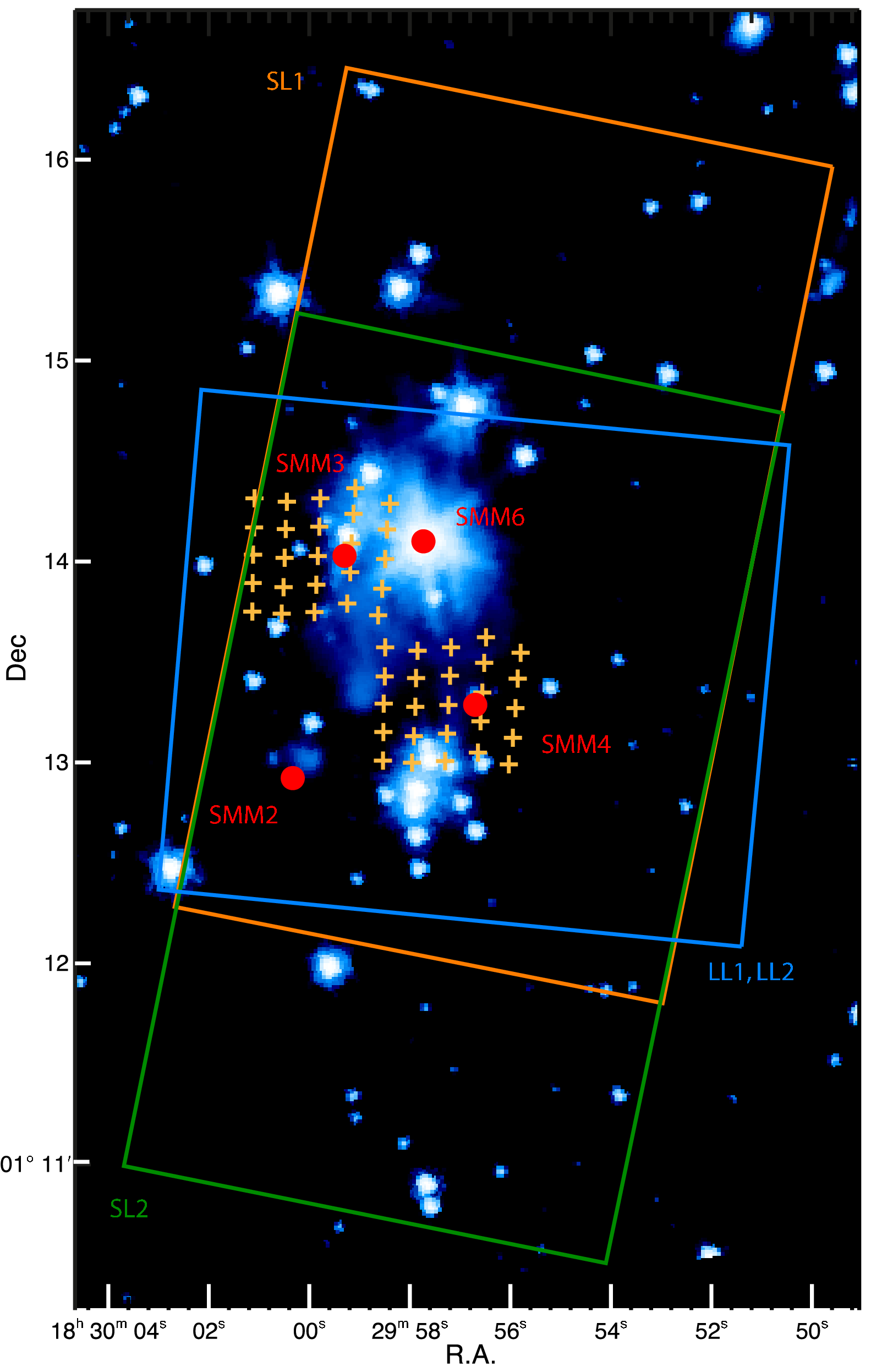}}
\caption{Regions of the Serpens SE cluster observed with Spitzer/IRS and
Herschel/PACS superimposed on a Spitzer IRAC 4.5 $\mu$m image. Dark orange,
green  and blue rectangles correspond to the areas mapped with Spitzer/IRS
modules SL1, SL2 and LL1/LL2, respectively.  Orange crosses denote the
5$\times$5 Herschel/PACS footprints around the protostellar sources SMM3 and
SMM4. Positions of known embedded sources in the region are indicated with
filled red circles.} \label{fig:1} \end{figure}

Since its first identification as an active star-forming region by
\citet{Strom:76a}, the Serpens cloud core has attracted much attention owing to
its remarkably high stellar density and star formation efficiency
\citep[e.g.][]{Enoch:07a} observed within its limited extent  of a few
arc-minutes ($\sim$0.2-0.4 pc for the distance estimates ranging between 260 pc
\citep{Straizys:03a} and 415$\pm$25 pc \citep{Dzib:10a}, the latter adopted in
this work). Among the stellar population of the Serpens complex, $\sim$ 30
embedded (Class 0 \& I) protostars have been identified \citep{Winston:07a,
Harvey:07a}, distributed within two main clumps at the northwest and southeast,
as opposed to a dispersed population of more evolved (Class II) protostars along
the cluster. The continuum and line emission from embedded sources in Serpens
has been studied with ISO/LWS and CVF instruments \citep{Larsson:00a,
Larsson:02a}. Nevertheless, the limited spatial resolution ($\sim$ 80$\arcsec$)
of the LWS was not sufficient to constrain and differentiate
the possible mechanisms
responsible for the excitation of the gas. In addition, the available spectral
resolution (R$\sim$ 200) and sensitivity allowed only the detection of the
strongest, often blended, emission lines.

Here we present combined spectro-imaging observations of the embedded protostars
SMM3 and SMM4, located in the SE region in Serpens with Spitzer/IRS and Herschel
PACS. The combined power of both instruments  provides an almost complete
wavelength coverage between 5 and 200 $\micron$. These observations cover almost
the same wavelength regime as the ISO observations presented in
\citet{Larsson:02a} ($\sim$ 5-190 $\micron$) at an average angular resolution of
$\sim$9.4$\arcsec$ and at spectral resolutions ranging from R=60-120 (IRS) to
R=1500-3000 (PACS). Compared with ISO observations, Herschel/PACS provides an
improvement by a factor of $\sim$5 in angular and $\sim$10 in spectral
resolution.

The paper is organized as follows: Sect.~\ref{sec:2} presents the observations
and describes the reduction of the data. Sect.~\ref{sec:3} discusses the
emission morphology revealed by line and continuum maps around SMM3 and SMM4.
The underlying excitation conditions are derived with analytical methods and
further discussed in comparison to shock models in Sect.~\ref{sec:4}. Finally,
Sect.~\ref{sec:5} puts into context the main results from the analysis and
Sect.~\ref{sec:6} provides a summary of this work.

\section{Observations and data reduction} \label{sec:2}

\subsection{Herschel/PACS} \label{sec:2.1} Observations were obtained with the
Photodetector Array Camera and Spectrometer \citep[PACS;][]{Poglitsch:10a} as
part of the ``Dust, Ice and Gas in Time (DIGIT)" open time key-project
\citep{Green:12a}. PACS is a 5$\times$5 array of 9.4\arcsec$\times$9.4$\arcsec$
spatial pixels (referred to as spaxels). The spectral range extends from 51 to
210 $\micron$ with R $\sim$ 1000--3000, divided into
four segments, covering $\lambda \sim$ 50--75, 70--105, 100--145, and 140--210
$\mu$m. The half power beam size of Herschel ranges from $\sim$5$\arcsec$ at 50
$\micron$ to $\sim$13$\arcsec$ at 200 $\micron$ and therefore the nominal spaxel
size of PACS (9.4$\arcsec$)  is a compromise between these two limits.

Observations were performed in range-scan spectroscopy mode, providing the
complete coverage of the wavelength range observable by PACS.  Two footprints
were observed targeting the protostellar sources SMM3 ($\alpha_{J2000}$=
18$^h$29$^m$59$^s$.3, $\delta_{J2000}$=+01$^d$14$^m$01$^m$.7) and SMM4
($\alpha_{J2000}$=18$^h$29$^m$56$^s$.7, $\delta_{J2000}$=+01$^d$13$^m$17$^m.2$),
as reported in \citet{Harvey:07a}.  Both sources were observed on April 2nd,
2010 with an integration time for each footprint of 4.2 hours. The telescope and
sky background emission was subtracted using two nod positions 6\arcmin\ from
the source in opposite directions. The observations were done in single staring
mode, that is the obtained maps are not Nyquist sampled. For both sources
observations were mispointed by $\sim$7$\arcsec$ to the west, as presented in
Fig.~\ref{fig:1}.  This offset is attributed to the instrument pointing
calibration accuracy during the first phases of the mission, which exhibited
residuals of this magnitude (see the Herschel Pointing Calibration Report, v1.0
- HERSCHEL-HSC-DOC-1515).
The pointing offset has been accounted for in all comparisons to other
data in the following sections.

 \begin{figure} \centering
\resizebox{\hsize}{!}{\includegraphics{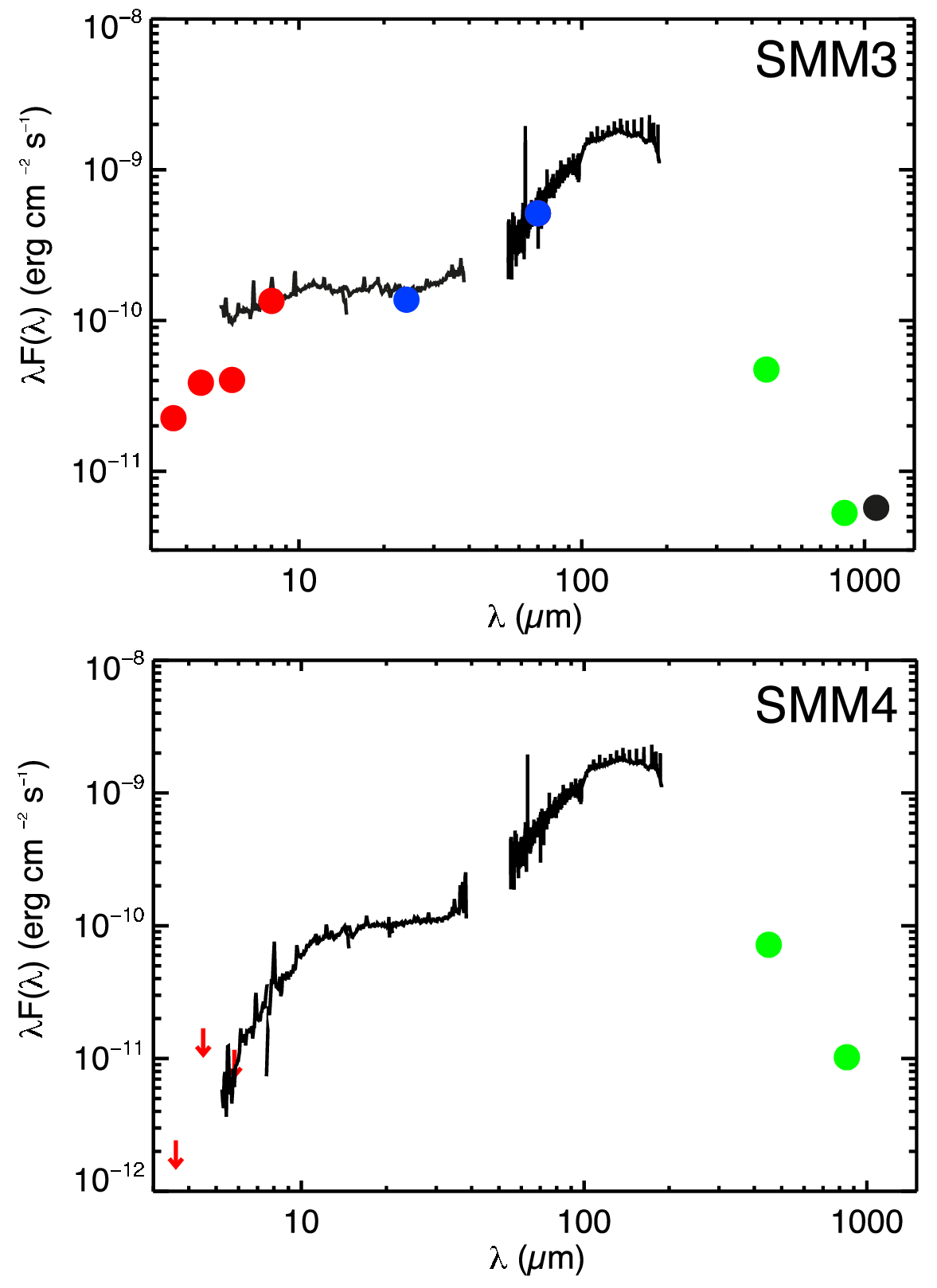}}
\caption{Spectral energy distribution diagram of SMM3 and SMM4 (upper and lower
panels, respectively) along with spectra from Spitzer/IRS and Herschel/PACS.
Fluxes for both instruments are extracted within a region encompassing the 3x3
central spaxels of PACS.  Filled circles are flux densities from IRAC at 3.6,
4.5. 5.8 and 8.0 $\micron$ bands \citep[red,][]{Harvey:07a}, MIPS at 24 and 70
$\micron$ \citep[blue,][]{Harvey:07a}, SCUBA at 450 and 850 $\micron$,
\citep[green,][]{Davis:99a}  and 1100 $\micron$ \citep[black,][]{Enoch:07a}.
Red arrows represent upper limit values for the IRAC bands.} \label{fig:2}
\end{figure}

\begin{figure} \centering
\resizebox{\hsize}{!}{\includegraphics{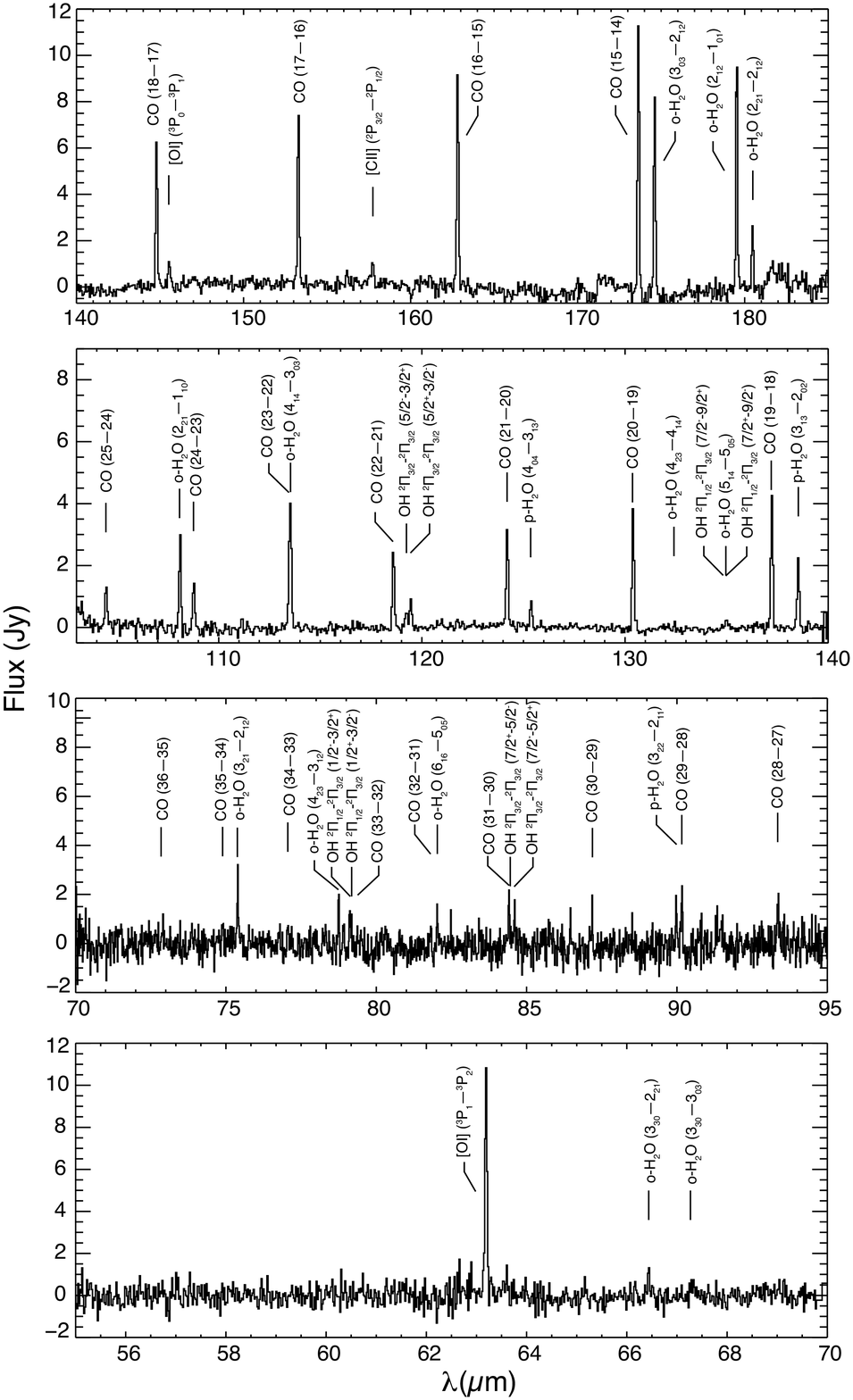}}
\caption{PACS continuum-subtracted spectrum from a single spaxel at the NW of
SMM3. Molecular and atomic transitions are reported on top of each detected
line.} \label{fig:3} \end{figure}

The data were reduced following the general procedure adopted for the DIGIT
embedded objects, described in detail in \citet{Green:12a}. Summarizing,
the reduction pipelines provide us with two data products which are based on
HIPE versions 6.1 and 8.0, respectively.  For the first data product (DP1) we
employ the telescope background calibration method (HIPE 6.1) which results in
best continuum matches between the different spectral segments. In addition,
absolute flux calibration is found to be good to a 10\% level when compared to
Spitzer and Herschel photometry (see Fig.~\ref{fig:2}). The second data product
(DP2) is based on the calibration block method (HIPE 8.0)  and provides a better
signal-to-noise ratio (SNR) on individual lines. In this dataset however the
overall continuum fluxes are not well calibrated and spectral segments show
discontinuous jumps. The continuum and line fluxes between the two datasets
scale by the same factor, which varies with wavelength. Therefore the flux
calibrations in DP2 are affected by a multiplicative calibration error.
Representative DP2 spectra (Fig.~\ref{fig:3}) display a significant number of
lines from CO, H$_2$O, OH, [OI] and [CII]. Spectra in the blue segments ($<$100
$\micron$) have an average noise level of $\sim$ 0.25 Jy, more than double
compared to the $\sim$ 0.1 Jy in the red modules. No other lines were identified
after thorough comparisons against molecular and atomic line catalogues.

Line fluxes were determined for each individual spaxel. They were initially
calculated in  DP2  by fitting the line profile with a gaussian, after having
subtracted a first order polynomial baseline. Subsequently, line fluxes were
scaled to match DP1 levels providing the correct flux levels.  Additional
corrections for flux spillover were applied by scaling the line fluxes with the
instrumental PSF. This assumes that most of the emission arises from selected individual spaxels and the spill over between these spaxels is not significant. Disentangling the origin of extended emission can be complex (see Sect.~\ref{sec:4.0}) and PSF corrections may underestimate the total emission from a single spaxel \citep{Karska:13a}. Spectra in the range between 96 and 103 $\micron$ are
exceedingly noisy and show irregular variations in the the continuum flux
levels. Abnormally low flux levels are also evident beyond 190 $\micron$ along
with ``ghost" spectral features due to leakage of emission from higher
orders. Therefore line fluxes from these two parts are unreliable and have
been excluded.

\subsection{Spitzer/IRS} \label{sec:2.2} Spitzer observations were retrieved
from the Spitzer Heritage Archive
(SHA)\footnote{http://sha.ipac.caltech.edu/applications/Spitzer/SHA/}.  They
have been performed as part of the c2d program \citep{Evans:03a, Evans:09a,
Lahuis:10a}.  In these data, the low resolution (R
$\sim$ 60 - 120) modules short-low (SL) and long-low (LL) of the Spitzer
Infrared Spectrograph \citep[IRS,][]{Houck:04a} were employed, providing a
complete wavelength coverage between 5.2 and 38.0 $\micron$.  Observations were
performed in slit-scan mode consisting of consecutive integrations after
shifting the slit to the parallel and perpendicular directions in relation to
the slit length, until the desired area is covered.  The SL scans consist of 5
$\times$ 43 such observations, where the SL slit was offset by 3.5$\arcsec$ in
the parallel and 50$\arcsec$ in the perpendicular directions, covering a total
area of 145$\arcsec \times $255$\arcsec$.  Similarly, the LL scans consist of
1x15 observations, offsetting the LL slit by 9.5$\arcsec$ only in the parallel
direction (Fig.~\ref{fig:1}).  Integration times per pointing were  28 and 60
seconds for the SL and LL modules, respectively.

\begin{figure} \centering
\resizebox{\hsize}{!}{\includegraphics{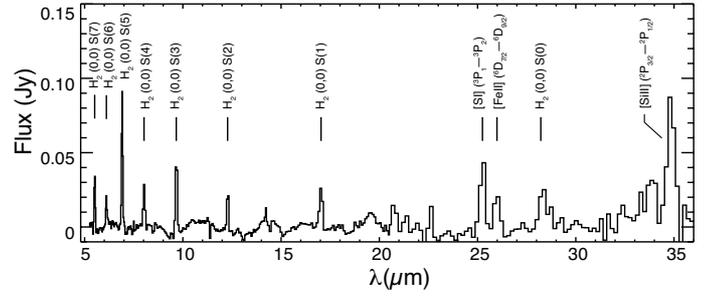}}
\caption{IRS spectrum extracted at the PACS spatial scale encompassing the same
region as  the PACS spectrum in fig \ref{fig:3}. Molecular (H$_2$) and forbidden
atomic transitions ([SI], [FeII], [SiII]) are marked on top of each line
detected. Unidentified line-like features are due to residual bad/rogue pixels}
\label{fig:4} \end{figure}

Initial data processing was performed with  version S18.7 of the
\textit{Spitzer} Science Center pipeline. Spectral data cubes were compiled
using the CUBISM software \citep{Smith:07a}, and bad/rogue pixels were masked by
visual inspection. As in the case of PACS, emission line maps were constructed
through customized procedures. In these, for each spaxel of a data-cube, the
flux for each spectral line of interest was calculated by fitting a Gaussian
after subtracting a local first or second order polynomial baseline. The
resulting line intensity maps for the IRS data have a square spaxel of side
equal to the width of the low resolution IRS modules (3.5$\arcsec$ and
10.5$\arcsec$ for the SL and LL modules, respectively), while the half power
beam size of Spitzer ranges between 3$\arcsec$ at 5.2 $\mu$m to 10$\arcsec$ at
38 $\mu$m. Full resolution line maps are presented in Fig.~\ref{fig:11} and in
appendix \ref{app:a}. For direct comparison with the PACS maps, IRS data cubes
were resampled according to the Herschel pointings at the PACS spaxel size of
9.4$\arcsec$, therefore providing analogous spectral line maps.  The
maximum half power beam size of Spitzer is comparable to the PACS spaxel
dimensions,
and therefore no significant flux losses are expected to occur after resampling
the IRS maps to the PACS grid.  
A resampled spectrum encompassing the same area
as PACS in Fig.~\ref{fig:1} towards the outflow of SMM3 is shown in Fig
\ref{fig:4}; strong emission lines from the first eight pure rotational
transitions of hydrogen (S(0) - S(7)), as well as atomic and ionic lines from
[FeII], [SI] and [SiII] are seen, indicating along with the PACS spectra highly
energetic conditions. 

\section{Spectral maps} \label{sec:3}

\subsection{Herschel} \label{sec:3.1}

\subsubsection{Continuum emission} \label{sec:3.1.3}

\begin{figure} \centering
\resizebox{\hsize}{!}{\includegraphics{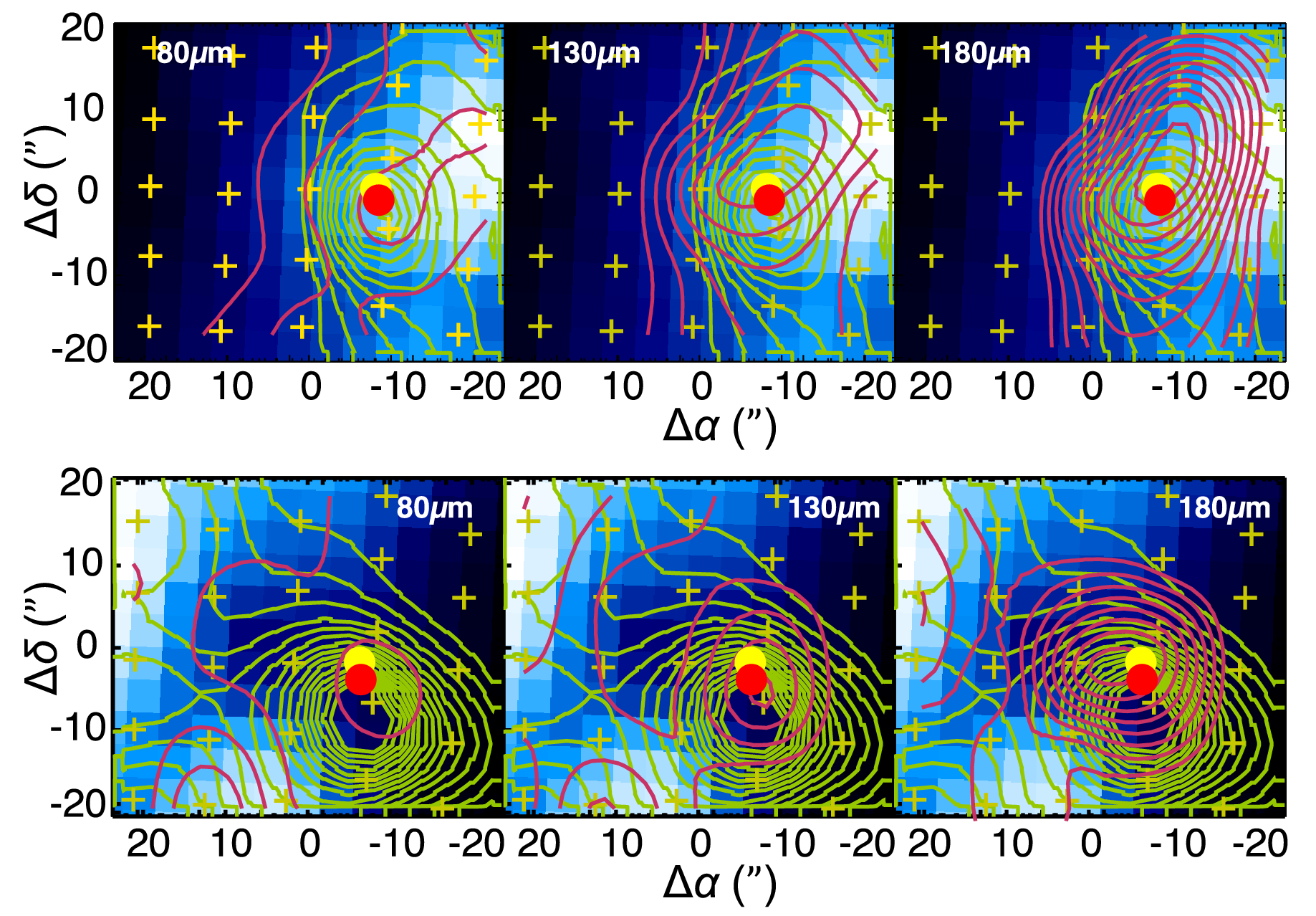}}
\caption{PACS continuum levels at 80, 130 and 180 $\mu$m (red contours)
superimposed on a Spitzer/MIPS 70 $\mu$m images around SMM3 and SMM4 (upper and
lower panels, respectively).  Green contours show the 450 $\mu$m continuum
observed with SCUBA, while the yellow and red dots depict the position of the
sources reported in \citet{Harvey:07a} and \citet{Hogerheijde:99a}. PACS and
SCUBA levels start at 2 Jy and increase by 2 Jy and 1.5 Jy steps, respectively.}
\label{fig:10} \end{figure}

Figure \ref{fig:10} presents maps of the  PACS continuum levels at 80, 130 and
180~$\micron$ (red contours, running from left to right) around SMM3 and SMM4
overlaid on top of a Spitzer/MIPS 70~$\micron$ image. In the same figure, green
contours shape the 450~$\micron$ continuum emission observed with SCUBA
\citep{Davis:99a}, while the filled yellow and red dots display the positions of
SMM3 and SMM4 derived from Spitzer \citep{Harvey:07a} and millimeter
interferometric observations \citep[at $\lambda=$3.4, 3.2, 2.7 and 1.4 mm,
][]{Hogerheijde:99a}.  PACS continuum emission at 80 $\micron$ follows well the
brightness distribution of the 70 $\micron$ underlying image, showing peaks to
the west of SMM3 and both to the NE and SE of SMM4. In the 130~$\micron$ and
180~$\micron$ maps, the continuum morphology gradually departs from the MIPS
brightness profile with the observed peaks shifting closer to the nominal source
positions.  The distribution of warm dust detected here is not symmetric around
the protostellar sources, indicating asymmetric envelope morphologies and
substantial continuum emission in the direction of outflows.
The emission at the longest continuum wavelength (180 \micron) from
PACS does not overlap completely with the SCUBA data at 450 \micron,
suggesting at peak farther south in column density than is apparent in
the PACS maps.

\subsubsection{Molecular emission} \label{sec:3.1.1}

\begin{figure} \centering
\resizebox{\hsize}{!}{\includegraphics{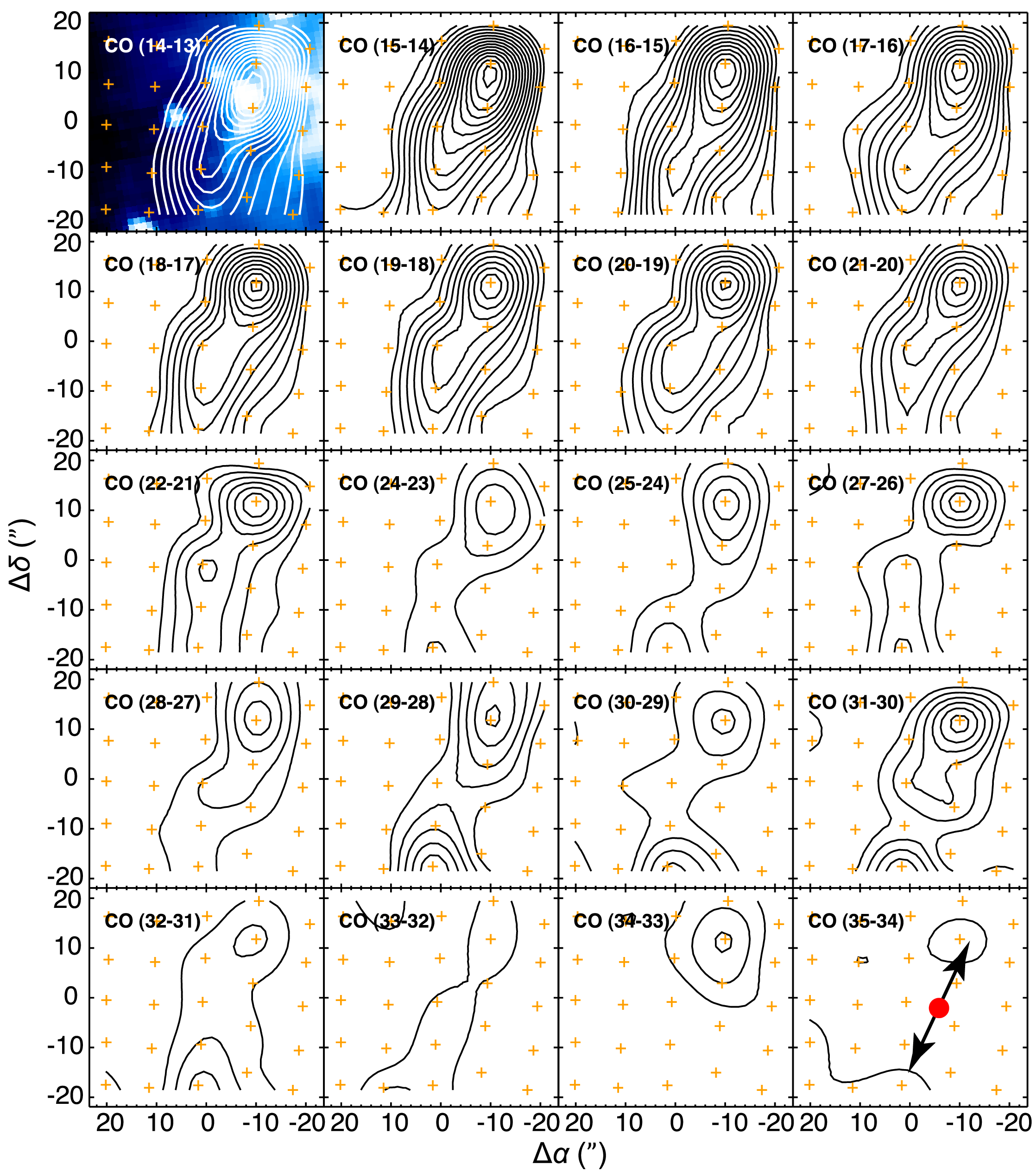}}
\caption{Spectral line maps of the $J$ = 14 -- 13 up to $J$ = 35 -- 34 CO
transitions around SMM3. All CO line maps presented show emission extending to
the NNW and SSE from the exciting source near the center of each panel, with a
strong peak towards the NW. The CO $J = $ 23 -- 22 transition is blended with
the stronger H$_2$O(4$_{14}$-3$_{03}$) line and is therefore presented in
Fig~\ref{fig:7}.  The enhancement in the $J =$ 31 -- 30 map is due to blending
with the OH (3/2-3/2, 7/2$^+$-5/2$^-$)  line.  A 4.5$\mu$m Spitzer/IRAC image is
presented as background on the upper left panel; bright regions correspond to
H$_2$ emission which is spatially coincident with the CO peak. The filled circle
and arrows in the lower-right panel display the position of the source
\citep{Harvey:07a} and the direction of the outflows \citep{Dionatos:10a}.
Contour levels are from 10$^{-14}$ erg cm$^{-2}$ s$^{-1}$ and increase at steps
of 10$^{-14}$ erg cm$^{-2}$ s$^{-1}$ ($\sim$3-sigma of the weakest transitions)}
\label{fig:5} \end{figure}

Figure~\ref{fig:5} presents integrated line emission maps of the CO transitions
detected with PACS around SMM3. CO emission is extended, exceeding the
dimensions of the PACS footprint in the N-S direction except for the transitions
with $J_{up} \ge 34$ where it shows a single peak. The emission pattern runs
roughly in the NNW-SSE direction, and strongly peaks towards the NW from the
center, at a distance of $\sim 20 \arcsec$. At roughly the same distance from
the center, a secondary peak becomes apparent for intermediate $J$ transitions
($24 < J_{up} < 31$) towards the southern edge of the map. At higher energy
transitions, emission becomes weaker and the emission pattern less clear. 

\begin{figure} \centering
\resizebox{\hsize}{!}{\includegraphics{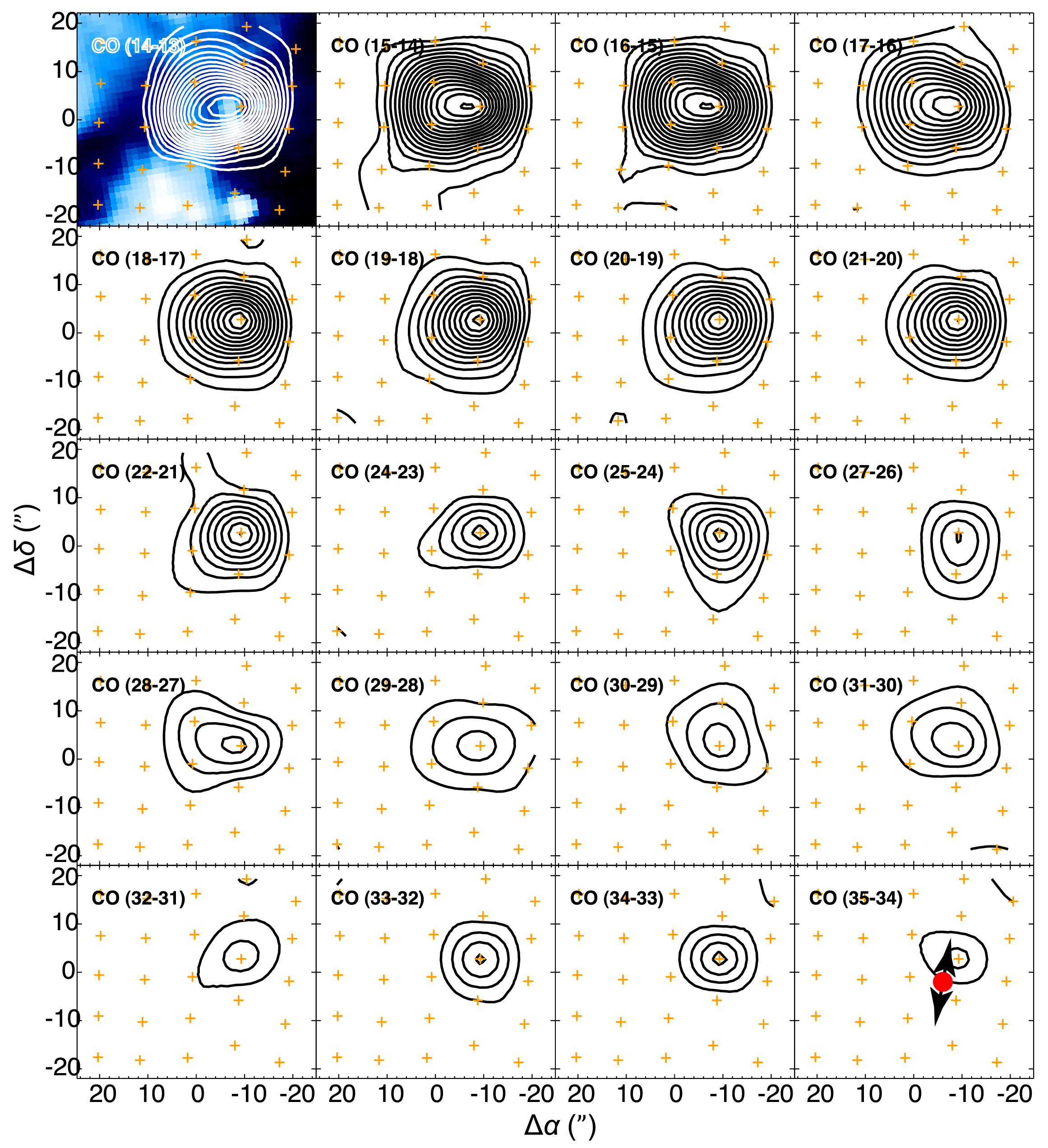}}
\caption{As in Fig.~\ref{fig:5} for SMM4. } \label{fig:6} \end{figure}

Maps of the CO lines detected with PACS around SMM4 are presented in
Fig.~\ref{fig:6}.  The line emission pattern is more centralized than in the
case of SMM3, and peaks off the central spaxel and to the north of the
protostellar source at a distance of $\sim5\arcsec$. Similarly to  SMM3,
emission becomes weaker with increasing $J$ showing only local maxima for
$J_{up} \ge 32$. Therefore the high-$J$ and mid-$J$ CO lines do not necessarily
co-exist in all positions.

\begin{figure} \centering
\resizebox{\hsize}{!}{\includegraphics{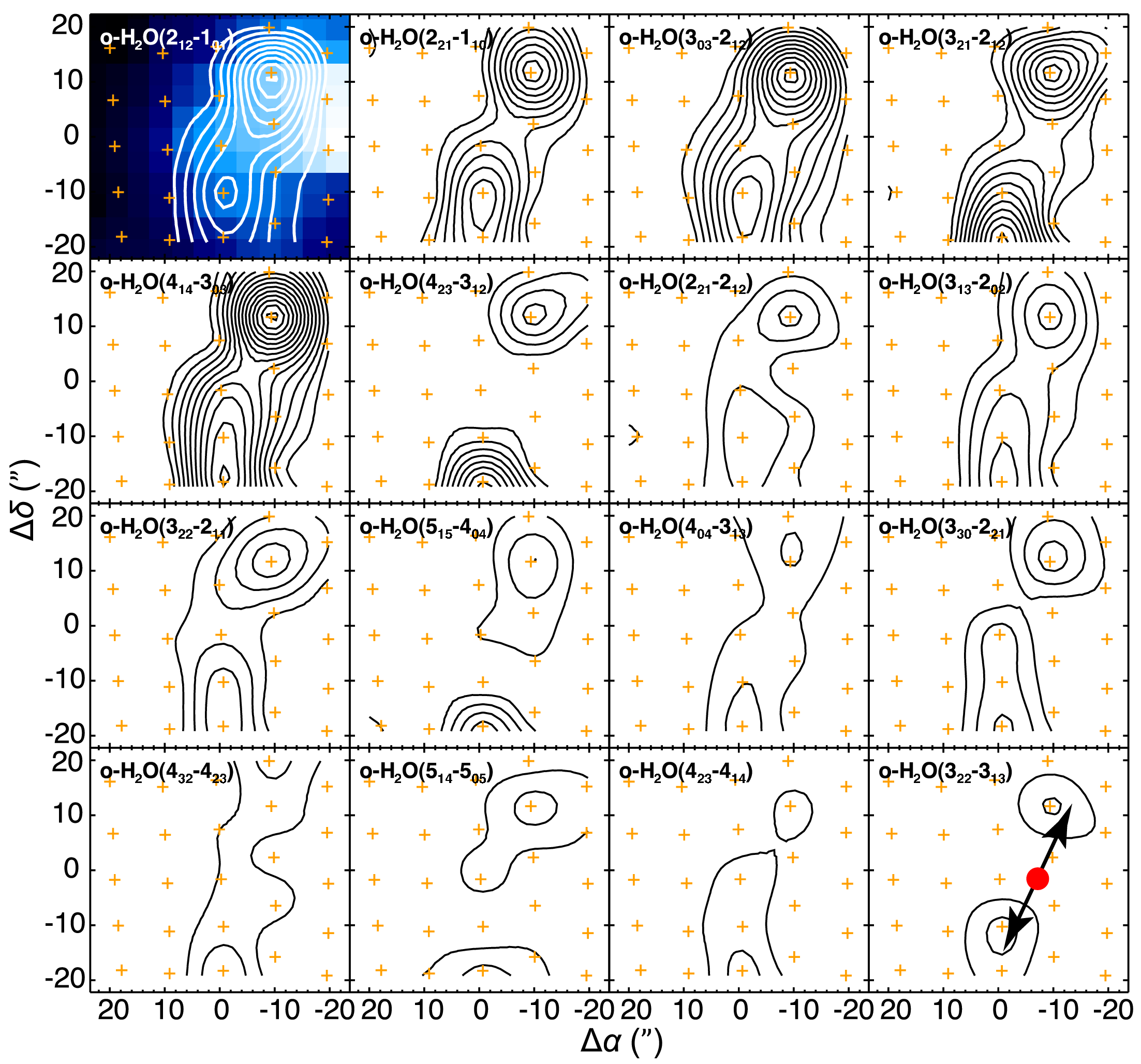}}
\caption{Spectral line maps of the strongest water transitions detected by PACS
towards SMM3. H$_2$O emission has very similar morphological characteristics to
the CO maps (see Fig.~\ref{fig:5}), with the only exception traced to the SSE
outflow from SMM3, where a secondary peak may be traced.  Contour levels are as
in Fig.~\ref{fig:5}.} \label{fig:7} \end{figure}

\begin{figure} \centering
\resizebox{\hsize}{!}{\includegraphics{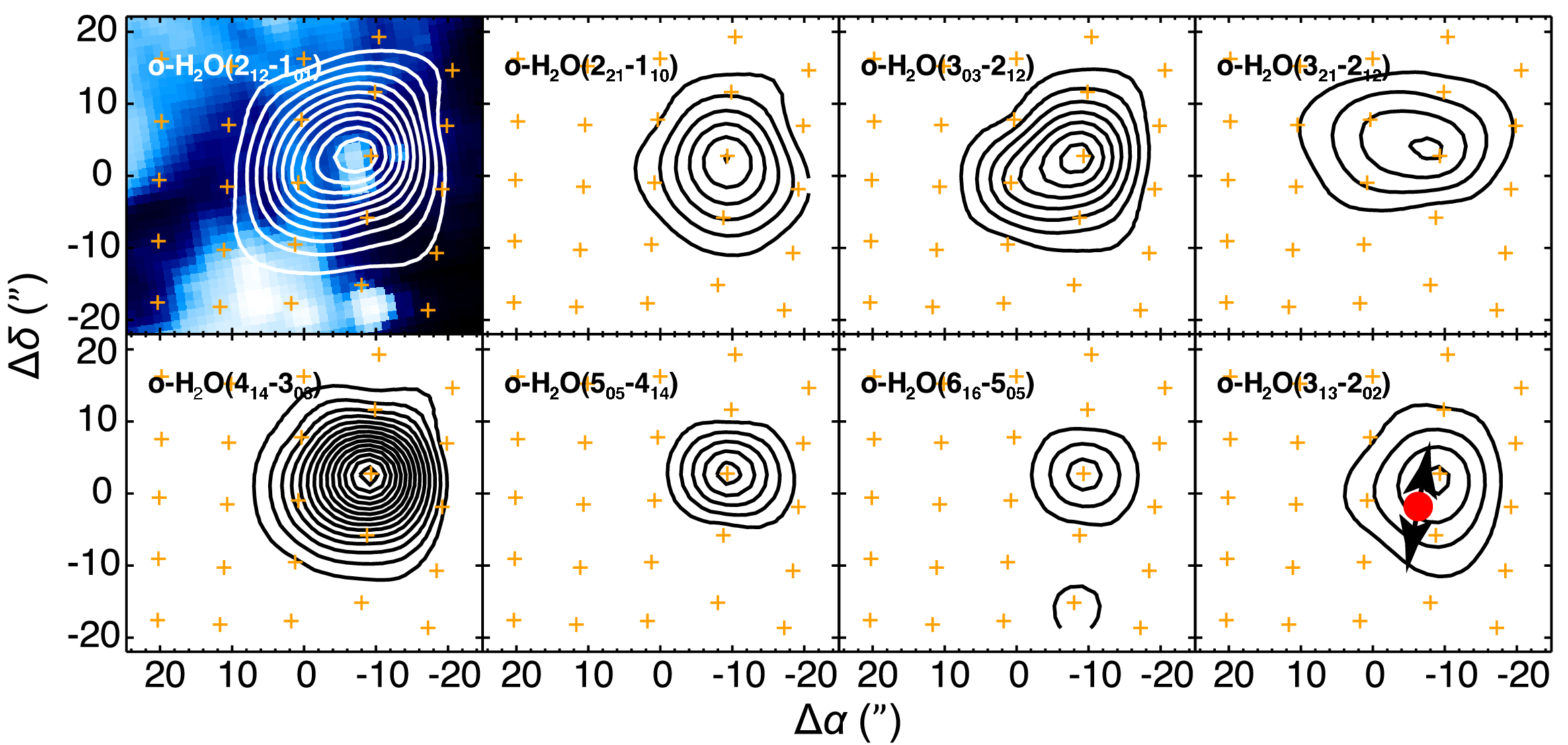}}
\caption{As in Fig.~\ref{fig:7} for SMM4.} \label{fig:8} \end{figure}

Figs.~\ref{fig:7} and \ref{fig:8} present the strongest water lines around SMM3
and SMM4 traced by PACS. For both sources, the observed morphologies are very
similar to the CO maps, with the only exception being a secondary peak observed
towards the southern lobe of SMM3 at $\sim10\arcsec$ from the center, in a
similar fashion to the high-$J$ CO lines. This difference in the emission
pattern of the two molecules may reflect differences in their excitation (see
Sect.~\ref{sec:4}).  H$_2$O line emission maps display a decline in intensity
moving from lower to higher excitation energy transitions, which eventually is
concentrated around the spaxels showing emission maxima.

\begin{figure} \centering
\resizebox{\hsize}{!}{\includegraphics{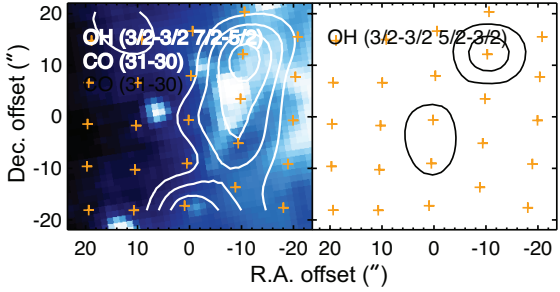}}
\caption{Extended OH emission around SMM3. Contour levels are as in
Fig.~\ref{fig:5}.} \label{fig:8b} \end{figure}

OH doublets at 71.2 and 79.1 $\micron$ are not spectrally resolved at the
signal-to-noise ratio in the blue segments, and resolved doublets at 84.4 and
134.8 $\micron$ are blended with much stronger CO and H$_2$O lines (see
Table~\ref{tab:1}). OH  emission (Fig.~\ref{fig:8b}) is in general weak
and lines are detected in most cases at the spaxels where CO and H$_2$O peak,
with the single exception of the OH ($^2\Pi_{\frac{3}{2}} -
^2\Pi_{\frac{3}{2}}$,  $J = \frac{5}{2}^{-} - \frac{3}{2} ^{+}$) line at 119.2
$\micron$ which appears to follow the extended emission pattern traced by CO and
H$_2$O.

The molecular emission morphology can provide indications of the underlying
processes responsible for the excitation of the gas. Despite small differences,
CO, H$_2$O and OH trace similar or related morphologies seen in each source.
However, the pattern observed around SMM3 is substantially different from that
seen around SMM4. Emission around the former source is in all cases extended,
whereas in the latter case, lines peak off the protostellar source, but the
emission pattern is more compact. In both cases, emission peaks offset from the
sources, but the extent and symmetry of the observed structures present no
further resemblance.

\begin{figure} \centering
\resizebox{\hsize}{!}{\includegraphics{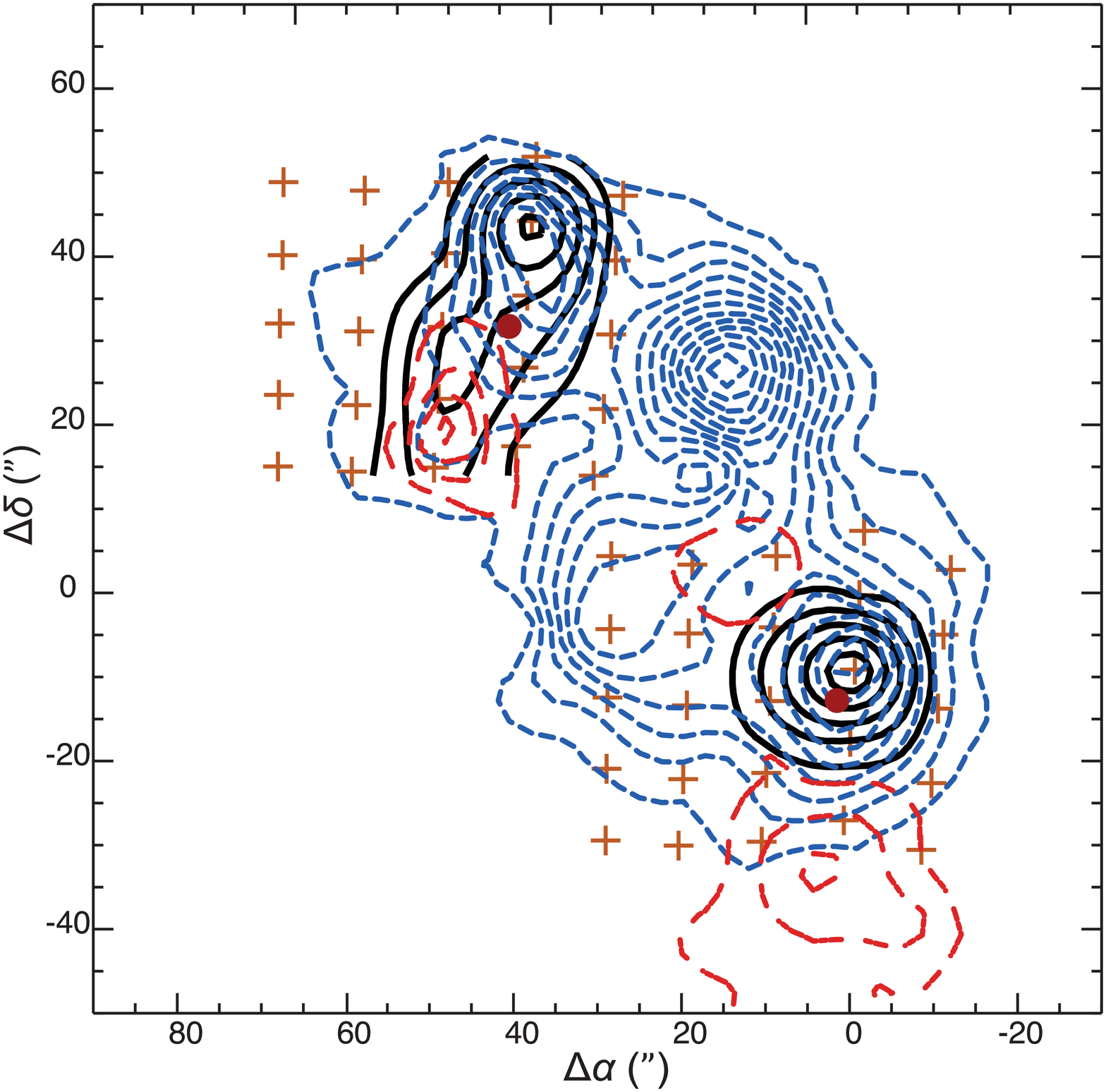}}
\caption{CO $J = 18 - 17$ emission observed with Herschel/PACS (black continuous
contours)  superimposed on high velocity CO $J = 3 - 2$ blue-shifted (blue,
short-dashed contours) and red-shifted (red, long-dashed contours) emission from
JCMT/HARP-B  \citep{Dionatos:10a}. Positions of the PACS spaxels are indicated
with (orange) crosses, and the locations of SMM3 and SMM4 with (red) filled
circles. Peaks of low and high $J$ CO emission are coincident.} \label{fig:8a}
\end{figure}

Observations of high velocity emission from low energy levels of 
CO, such as  $J = 3-2$
%\citep[e.g. $J = 3-2$]
%[\textbf{see also Fig.} \ref{fig:8a}]{Dionatos:10a, Graves:10a} 
\citep{Dionatos:10a, Graves:10a} and see Fig. \ref{fig:8a},
and $J = 2-1$, \citep{Davis:99a}, at resolutions $\sim$7$\arcsec$--15$\arcsec$,
comparable to those provided by PACS, show very similar structures which are
attributed for both sources to outflows. In the case of SMM3, the NNW and SSE
lobes appear to be blue- and red-shifted, respectively, whereas in the case of
SMM4, the emission structure coincides with blue-shifted gas. A red-shifted lobe
to the SE, often attributed to SMM4 \citep[e.g.][]{Narayanan:02a, Dionatos:10a}
is not traced here by PACS.

Around both SMM3 and SMM4, shock excited H$_2$ emission following the outflow
pattern has been recorded in the near and mid-IR \citep[e.g.][see also the
following section \ref{sec:3.2}]{Herbst:97a, Eiroa:97a, Larsson:02a}. Similar
outflows are recorded in the methanol maps of \citet{Kristensen:10b}. 
High resolution maps in several molecular lines,
such as HCO$^+$, HCN, SiO \citep{Hogerheijde:99a} and CS \citep{Testi:00a}
reveal narrow, jet-like structures following the outflow orientations.

\subsubsection{Atomic emission} \label{sec:3.1.2}

\begin{figure} \centering
\resizebox{\hsize}{!}{\includegraphics{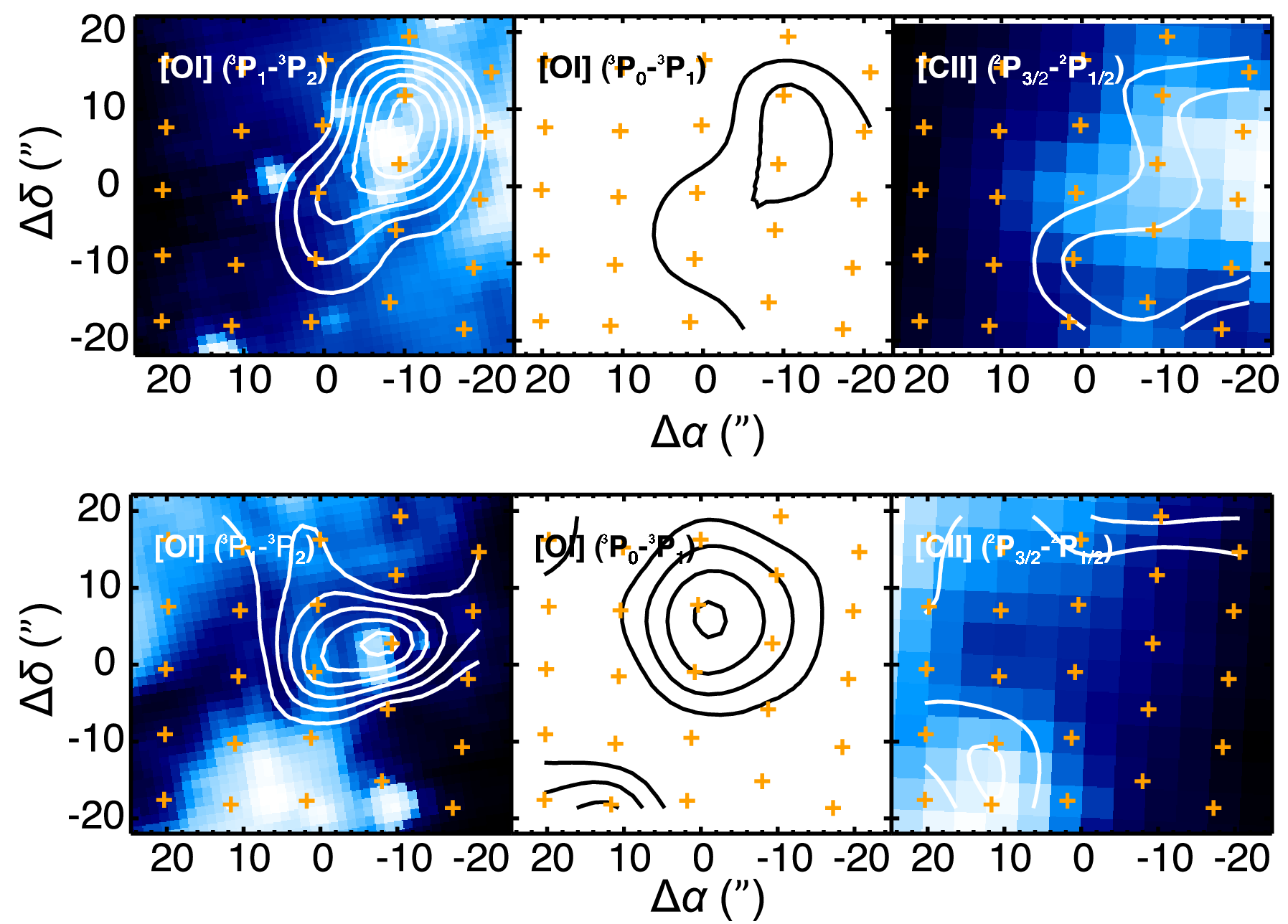}}
\caption{Spectral line maps for the atomic lines detected with PACS around SMM3
and SMM4 (upper and lower panels, respectively).  Oxygen lines follow the
outflow pattern shown in molecular line maps while [CII] cannot correlate with
the outflow morphology. Contour levels are as in Fig.~\ref{fig:5} except for the
[OI] $^3$P$_1$ - $^3$P$_2$ maps which are from 10$^{-14}$ erg cm$^{-2}$ s$^{-1}$
with increments of 5$\times$10$^{-14}$ erg cm$^{-2}$ s${-1}$. Background images
are 4.5 $\micron$ from Spitzer/IRAC (left panels) and 70 $\micron$ Spitzer/MIPS
(right panels), display associations with the corresponding superimposed atomic
lines.} \label{fig:9} \end{figure}

Line maps of atomic species observed with PACS around SMM3 and SMM4 are
presented in Fig.~\ref{fig:9} (upper and lower panels, respectively). The
emission pattern for the [OI] line at 63.2 and 145 $\micron$ around both
sources is similar, given the lower signal-to-noise and resolution of the latter
line.  Both lines follow the pattern delineated by the molecular tracers and
peak at the same positions.
 
The [CII] emission line maps reveal a very different morphology. Around SMM3,
[CII] becomes most prominent at the western edge of the mapped region, whereas
for SMM4 it peaks to the north and SE of the PACS map. For both sources, the
[CII] emission appears not associated with the outflow patterns traced by the
molecular lines and the [OI] line at 63.2 $\micron$. It may only be associated
with the extended [OI] line pattern at 145.5 $\micron$, to the west and SE edges
of the maps around SMM3 and SMM4.  [CII] is a common tracer of photon-dominated
regions \citep[PDRs, e.g.][]{Dedes:10a} and as such, it responds to a source of
UV radiation.  Indeed, the emission west of SMM3 and north of SMM4 is consistent
with a common UV source in the NW quadrant between the two PACS footprints (see
Fig.~\ref{fig:1}). A possible candidate is SMM6, a binary, flat-spectrum source
\citep{Haisch:02a, Winston:07a} associated with energetic emission in Br$\gamma$
and X-rays \citep{Winston:07a, Preibisch:03a}. The bolometric luminosity of SMM6
($\sim$13 L$_{\odot}$) is more than 10 times higher than SMM3 and SMM4
\citep{Dionatos:10b}.   The PDR excitation of [CII] is further supported by the
corresponding map recorded with ISO/LWS \citep[upper right panel in their Fig. 3
of][]{Larsson:02a}, which covers a larger area than the PACS observations in
this paper: the kidney-shaped emission-maxima in [CII] for those maps extending
from SMM6 to the SE towards SMM3 and SMM4 is consistent with the former being
the exciting source. The [CII] emission mapped to the south of SMM4, is likely
to be associated with a cluster of more evolved protostellar sources at the south
\citep{Winston:07a, Harvey:07a}. In addition, the observed continuum morphology
in Sect. \ref{sec:3.1.3}, is consistent with the hypothesis of PDR excited [CII]
emission, as the material becomes denser and more effectively shielded towards
the protostellar sources.  The apparent resemblance between the morphology of
the [CII] maps and the 70$\micron$ Spitzer image (right panel of
Fig.~\ref{fig:9}) can be attributed to light recorded in the wide-band Spitzer
filter originating from the same exciting source heating the ambient dust.

\begin{table*}[!ht] \caption{\textit{Herschel} - PACS line fluxes measured at
selected positions of peak emission. Point-spread-function (PSF) corrections have
been applied to the reported levels and  errors are at 3-sigma level}
\label{tab:1} {\centering \begin{tabular}{l c c c c c c c } \hline\hline Element
&  Transition & Wavelength ($\mu$m)   &      \multicolumn{4}{c}{Flux (10
$^{-14}$ erg cm$^{-2}$ s$^{-1}$)} \\ &   &      &   SMM3  b &  SMM3 c   & SMM3 r
&   SMM4 b \\

\hline \hline

  $[$OI$]$ &           $^3$P$_1 - ^3$P$_2$    		& 63.1837         & 58.85$\pm$4.34      			&        29.04$\pm$3.48	     	& 22.04$\pm$3.14 &    52.31$\pm$3.52    \\ 
o-H$_2$O      &           $3_{30} - 2_{21}$     		       & 66.4372         & 4.25$\pm$1.35 & 3.15$\pm$1.28    		&         3.11$\pm$1.37 & 1.90$\pm$1.54   \\
o-H$_2$O      &           $3_{30} - 3 _{03}$ & 67.2689         & 6.17$\pm$2.42 &            $\ldots$ &         8.25$\pm$1.74 &     4.25$\pm$1.74   \\ 
CO		&           $38 - 37$ & 69.0744         &          $\ldots$ &	$\ldots$ &               $\ldots$    		& 2.50$\pm$1.58  \\ 
CO       & $37 - 36$        						& 70.9072 &	2.45$\pm$1.31      			&         $\ldots$ &         	$\ldots$	& 1.78$\pm$1.07   \\ 
OH$^a$ &  $^2\Pi_{\frac{1}{2}} - ^2\Pi_{\frac{1}{2}}$,$\frac{7}{2}^{-} - \frac{5}{2} ^{+}$  			& 71.1708         & \multirow{2}{*}{2.62$\pm$0.24} & \multirow{2}{*}{$\ldots$} 	& \multirow{2}{*}{$\ldots$}	& \multirow{2}{*}{3.05$\pm$0.45}   \\ 
OH$^a$ &   $^2\Pi_{\frac{1}{2}} - ^2\Pi_{\frac{1}{2}}$, $\frac{7}{2}^{+} - \frac{5}{2}^{-}$ & 71.2158         & \\ 
 o-H$_2$O      &           $7_{07} - 6_{16}$ & 71.9460         & $\ldots$           				& $\ldots$ &         1.75$\pm$0.84    	& 2.19$\pm$1.52    \\ 
CO       &           $36 - 35$ & 72.8429         & 4.35$\pm$1.08      			& $\ldots$ &         $\ldots$		    	& 2.95$\pm$1.40   \\ 
CO       & 		$35 - 34$        						& 74.8901 &	2.57$\pm$1.62      			&            $\ldots$ &	$\ldots$ &           3.38$\pm$1.72    \\ 
o-H$_2$O      & $3_{21} - 2_{12}$	& 75.3804         & 14.07$\pm$2.04 &         1.92$\pm$0.71    		&	7.74$\pm$1.61    	& 5.79$\pm$2.01   \\ 
CO       &           $34 - 33$ &	77.0587         & 4.52$\pm$2.82      			&            $\ldots$ &	$\ldots$ &     3.29$\pm$1.57   \\ 	
o-H$_2$O      & $4_{23} - 3_{12}$	& 78.7414         & 8.94$\pm$1.74 &            $\ldots$        		&	3.74$\pm$0.40 & 3.05$\pm$1.75   \\ 
OH$^a$       &    $^2\Pi_{\frac{1}{2}} -^2\Pi_{\frac{3}{2}}$, $\frac{1}{2}^{-} - \frac{3}{2} ^{+}$ & 79.1173         &\multirow{2}{*}{6.35$\pm$1.91 }    & \multirow{2}{*}{2.10$\pm$0.70}	&\multirow{2}{*}{$\ldots$  }  & \multirow{2}{*}  { 4.44$\pm$0.54  } \\ 
OH$^a$ &   $^2\Pi_{\frac{1}{2}} - ^2\Pi_{\frac{3}{2}}$, $\frac{1}{2}^{+} - \frac{3}{2}^{-}$ & 79.1809         &         \\ 
CO       &           $33 - 32$ & 79.3598 & 2.28$\pm$1.92      			&         1.62$\pm$1.40 & 2.37$\pm$1.42 &           2.58$\pm$1.44    \\ 
CO       &           $32 - 31$         						& 81.8058         &  3.31$\pm$2.78      			&         2.34$\pm$1.27    	& 2.99$\pm$1.90    	& 3.72$\pm$1.72   \\ 
o-H$_2$O      &           $6_{16} - 5_{05}$    			& 82.0304         & 5.17$\pm$1.31 & $\ldots$		    		&         4.04$\pm$0.75 & 5.09$\pm$1.51 \\ 
CO$^a$       &           $31 - 30$ & 84.4107        & \multirow{2}{*}{9.55$\pm$1.75} & \multirow{2}{*}{4.13$\pm$0.63} &\multirow{2}{*} {1.73$\pm$1.44} & \multirow{2}{*}  {5.53$\pm$1.65}   \\ 
OH$^a$       &    $^2\Pi_{\frac{3}{2}} -^2\Pi_{\frac{3}{2}}$, $\frac{7}{2}^{+} - \frac{5}{2} ^{-}$ & 84.4199         & \\ 
OH       &   	 $^2\Pi_{\frac{3}{2}} - ^2\Pi_{\frac{3}{2}}$,$\frac{7}{2}^{-} - \frac{5}{2} ^{+}$ & 84.5963         & 4.30$\pm$0.76&         2.55$\pm$0.34 &         2.26$\pm$0.72 &     2.36$\pm$0.72    \\ 
CO 	   &           $30 - 29$ & 87.1904         & 4.86$\pm$0.43	&         2.50$\pm$0.72 &         5.17$\pm$2.75 &     5.23$\pm$0.76    \\
p-H$_2$O      & $3_{22} - 2_{11}$     			& 89.9878         & 7.36$\pm$0.81 &            2.53$\pm$1.14       	&         1.60$\pm$0.66 & 1.82$\pm$0.34    \\ 
CO       &           $29 - 28$ & 90.1630         & 7.59$\pm$1.60     			&         2.31$\pm$1.40 & 5.23$\pm$0.76    	& 3.10$\pm$0.30    \\ 
CO       &           $28 - 27$ & 93.3491         & 6.92$\pm$2.27      			&         3.08$\pm$1.67 &         2.22$\pm$0.86 &     6.38$\pm$1.16    \\ 
CO       &           $27 - 26$ & 96.7725         & 8.50$\pm$1.05     			&         3.48$\pm$0.72 &         3.52$\pm$0.80    	& 5.94$\pm$1.50    \\ 
CO       &           $25 - 24$        						& 104.445         & 7.01$\pm$0.98      			&         2.48$\pm$1.15 & 2.19$\pm$1.10 &     9.63$\pm$1.18   \\ 
o-H$_2$O      & $2_{21} - 1_{10}$ & 108.073         & 13.84$\pm$2.18 &         5.24$\pm$1.13    		& 10.93$\pm$1.48 & 9.15$\pm$2.31   \\ 
CO       &           $24 - 23$ & 108.763 & 7.96$\pm$1.78      			&         2.66$\pm$0.39 & 1.81$\pm$0.33 &     8.25$\pm$1.18   \\ 
CO$^a$       &           $23 - 22$ &113.458         & \multirow{2}{*}{24.84$\pm$1.07} &\multirow{2}{*}{13.60$\pm$1.12}& \multirow{2}{*}  {15.73$\pm$0.68}    & \multirow{2}{*}  {22.95$\pm$0.56} \\ 
o-H$_2$O$^a$      &           $4_{14} - 3_{03}$     		& 113.537 & \\ 
CO       &           $22 - 21$ & 118.581 & 11.01$\pm$0.79      			&         6.60$\pm$0.31 & 5.30$\pm$0.34 &     12.79$\pm$1.31    \\ 
OH       & $^2\Pi_{\frac{3}{2}} -^2\Pi_{\frac{3}{2}}$, $\frac{5}{2}^{-} - \frac{3}{2} ^{+}$ & 119.232         & 2.75$\pm$0.72 &         6.91$\pm$0.38    		& 2.45$\pm$0.37 & 5.96$\pm$0.48   \\ 
OH       &  	$^2\Pi_{\frac{3}{2}} - ^2\Pi_{\frac{3}{2}}$, $\frac{5}{2}^{+} - \frac{3}{2} ^{-}$ & 119.440         & 3.75$\pm$0.40      			&            $\ldots$ & $\ldots$ &     6.56$\pm$0.50   \\ 
o-H$_2$O    & $4_{32} - 4_{23}$ & 121.719         & 1.30$\pm$.29          		&            $\ldots$ & 1.29$\pm$0.48 &           $\ldots$   	\\ 
CO       &           $21 - 20$ & 124.193         & 14.06$\pm$0.49      			&         8.44$\pm$0.55 &         7.22$\pm$0.42 &     16.44$\pm$0.37   \\ 
p-H$_2$O      & $4_{04} - 3_{13}$     			& 125.353         & 3.75$\pm$0.45 & 2.29$\pm$0.31   		&         3.57$\pm$0.55    	& 3.06$\pm$0.54 \\ 
CO       &           $20 - 19$ & 130.369         & 15.83$\pm$0.35 &         9.01$\pm$0.40 &         8.88$\pm$0.35 &    18.49$\pm$0.30   \\
o-H$_2$O      & $4_{23} - 4_{14}$     			& 132.407         & 1.27$\pm$0.25 &         1.74$\pm$0.35   		&         1.61$\pm$0.50 & $\ldots$    \\ 
OH$^a$       &  	$^2\Pi_{\frac{1}{2}} -^2\Pi_{\frac{3}{2}}$, $\frac{7}{2}^{-} - \frac{9}{2} ^{+}$		& 134.845         & \multirow{3}{*}  {2.37$\pm$0.65}   & \multirow{3}{*} {$\ldots$}      & \multirow{3}{*}  {$\ldots$}    &           \multirow{3}{*} {2.48$\pm$1.15}   \\ 
o-H$_2$O$^a$      &           $5_{14} - 5_{05}$ & 134.935 & \\ 
OH$^a$       &   	$^2\Pi_{\frac{1}{2}} -^2\Pi_{\frac{3}{2}}$, $\frac{7}{2}^{+} - \frac{9}{2} ^{-}$		& 134.964         & \\ 
CO       &           $19 - 18$ & 137.196         & 17.13$\pm$0.49      			&        10.70$\pm$0.34 & 10.37$\pm$0.31    & 21.48$\pm$0.39   \\ 
p-H$_2$O      & $3_{13} - 2_{02}$ & 138.527         & 8.48$\pm$0.43 &         5.41$\pm$0.49    		& 7.66$\pm$0.29    & 7.43$\pm$0.33   \\ 
CO       &           $18 - 17$ & 144.784 & 21.14$\pm$0.62      			& 11.64$\pm$0.48    		& 12.35$\pm$0.40 &     23.85$\pm$0.88 \\ 
$[$OI$]$ &       $^3$P$_0 - ^3$P$_1$ & 145.525         & 4.09$\pm$0.53      			&         2.83$\pm$0.44 &         3.22$\pm$0.25   & 4.74$\pm$3.37    \\ 
CO       &           $17 - 16$ & 153.267         & 22.19$\pm$0.73      			& 10.03$\pm$0.69 &         13.38$\pm$0.40 &    22.05$\pm$0.63   \\ 
p-H$_2$O      & $3_{22} - 3_{13}$     			& 156.193         & $\ldots$ & $\ldots$        		&         2.17$\pm$0.90 &    2.29$\pm$0.80 \\
$[$CII$]$  &           $^2$P$_{3/2} - ^2$P$_{1/2}$   	& 157.741 & 3.45$\pm$0.52      			&         1.01$\pm$0.49       	& 4.52$\pm$1.54 &     3.29$\pm$0.37    \\ 
CO       &           $16 - 15$ & 162.812 & 26.34$\pm$0.63      			& 12.12$\pm$0.73    		& 16.97$\pm$1.06 &    29.81$\pm$0.59   \\ 
CO       &           $15 - 14$ & 173.631 & 32.60$\pm$1.19      			& 18.10$\pm$0.58    		& 17.78$\pm$0.76 &    33.71$\pm$0.93   \\ 
o-H$_2$O      &           $3_{03} - 2_{12}$     			& 174.626         & 16.76$\pm$1.08 &         6.89$\pm$0.69 &         9.52$\pm$0.73    & 16.45$\pm$0.63   \\ 
o-H$_2$O & $2_{12} - 1_{01}$     			& 179.527         & 26.61$\pm$0.59 &         11.15$\pm$0.79    		&         16.81$\pm$0.90 & 23.93$\pm$0.65   \\ 
o-H$_2$O      &           $2_{21} - 2_{12}$ & 180.488 & 7.54$\pm$0.68      			&         4.65$\pm$1.40 & 6.02$\pm$0.54    & 5.09$\pm$0.59   \\ 
CO       &           $14 - 13$ & 185.999 & 30.53$\pm$0.90      			& 17.04$\pm$0.41    		& 19.11$\pm$0.74 &    37.83$\pm$0.62   \\

\hline \end{tabular} \\ } % This is for the centering of the table {}~$^a$
blended lines\\ \end{table*}

\subsubsection{Line emission pattern} \label{sec:4.0}

\begin{figure} \centering
\resizebox{\hsize}{!}{\includegraphics{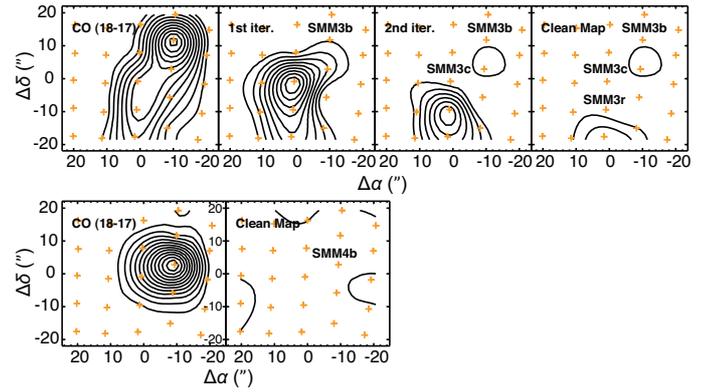}}
\caption{Deconvolution of the CO (18-17) maps around SMM3 and SMM4 with the
instrumental point-spread function.  The leftmost panel presents the
initial line map, and stepwise to the right each presents the residual map after
subtracting the instrumental PSF at a specific location. For SMM3 (upper
panels), such locations are indicated on each panel (SMM3b, SMM3c and SMM3r). In
SMM4 (lower panel) the emission is confined to a single spaxel (SMM4b). The
rightmost panel corresponds to the final residual map where flux levels are at
$\sim 10\%$ or lower compared to the peak values in the original maps.}
\label{fig:14} \end{figure}

The morphology of the line emission presented in the maps of the previous
section shows different characteristics for SMM3 and SMM4. To quantify whether
the emission is point-like or extended compared to the nominal PACS spaxel size
(9.4$\arcsec$), we employ the POMAC code \citep[][see also
Appendix~\ref{app:b}]{Lindberg:12a}, which performs a deconvolution of the
observed emission pattern with the instrumental point-spread function (PSF).
The code is based on a modified version of the CLEAN algorithm
\citep{Hogbom:74a} with the difference that it requires the positions of testing
``point'' sources. Such points in the present case are selected to be the
spaxels displaying emission maxima. The code was run iteratively on the
resulting residual maps of the previous clean process in the same fashion, until
reaching residual maps showing no significant emission. The results from this
analysis are presented in Fig.~\ref{fig:14} for the CO (18-17) line maps around
SMM3 and SMM4.  

The results of the cleaning process suggest that most of the line emission
around SMM3 is confined within 3 spaxels, towards the NW (blue peak), the center
and to the S (red peak), which are named hereafter SMM3b, SMM3c and SMM3r.
Excluding any of these points from the cleaning process results in
significant residuals at the location omitted, indicating that the emission
arises from unresolved regions within each spaxel.  In the case of SMM4 most
the observed emission originates from a single spaxel coinciding with the
blue-shifted lobe and named SMM4b (see also Fig.~\ref{fig:14}).  Line fluxes for
the molecular and atomic species observed with PACS are reported for these
points in Table~\ref{tab:1}. The following sections focus on the analysis of
individual spaxels indicated from the clean process.

\subsection{Spitzer} \label{sec:3.2}

\begin{figure} \centering
\resizebox{\hsize}{!}{\includegraphics{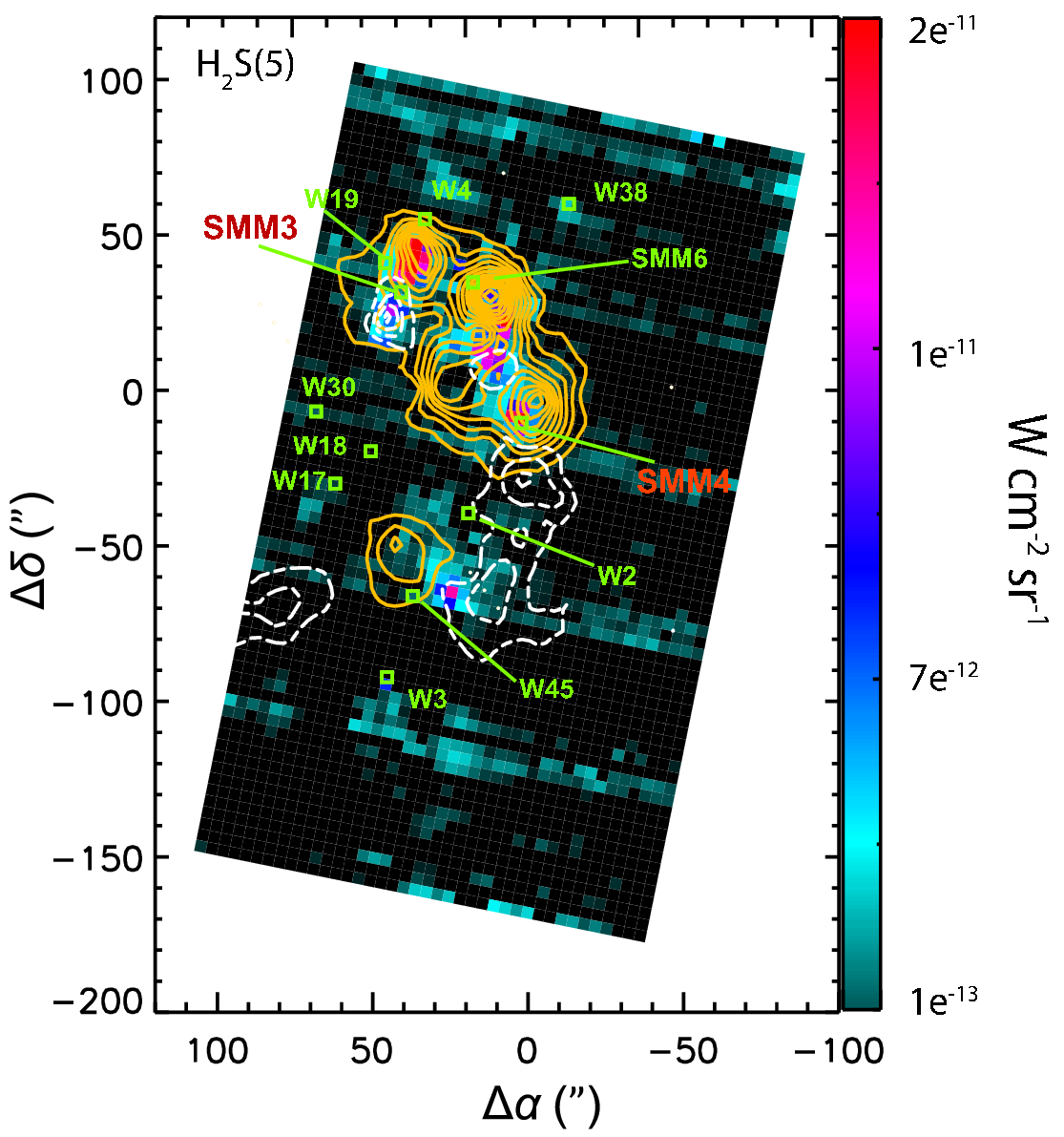}}
\caption{Spitzer spectral map of the Serpens SE region in H$_2$ 0-0 S(5) line
emission; color coded flux levels are indicated on the side bar. Dim stripes in
the horizontal direction are residuals due to rogue pixel contamination of the
IRS detectors. Yellow and white contours delineate high velocity blue- and
red-shifted $J = 3 - 2$ CO gas from \citet{Dionatos:10a}, showing a strong
spatial correlation in the excitation between the two species. Positions of
known protostellar sources from the catalogue of \citet{Winston:07a} are
indicated on the map.} \label{fig:11} \end{figure}

Spitzer/IRS maps encompass the whole SE region of the Serpens Cloud Core, at
resolutions ranging between 3.5$\arcsec$ to 10.5$\arcsec$ for the SL and LL
modules, respectively (see also Sect. \ref{sec:2}). A line intensity map at
3.5$\arcsec$ resolution  of the 0-0 S(5) pure rotational transition of molecular
hydrogen is presented in Figure \ref{fig:11}. Superimposed on this, yellow and
white contours corresponding to high velocity CO $J=3 - 2$ blue- and red-shifted
outflow gas from \citep{Dionatos:10a} show a clear spatial association between
H$_2$ and entrained gas traced by low-$J$ CO. In the same figure, the positions
of protostellar sources extracted from the catalogue of \citet{Winston:07a} are
also indicated. The close association between low-$J$ CO and H$_2$ emission is
also seen in the outflows of L1157 \citep{Nisini:10a}.  Serpens maps with
ISO/CVF \citep{Larsson:02a} at 7~$\mu$m show substantial emission around the SW
clump, with significant enhancements to the north of SMM3 and close to SMM6,
which are in good correspondence with the morphology shown in Fig.~\ref{fig:11}.
The full Spitzer emission line maps for all the detected H$_2$ and forbidden
atomic transitions from [FeII], [SiII] and [SI] are presented in Appendix
\ref{app:a}.

\subsubsection{H$_2$ emission} \label{sec:3.2.1}

\begin{figure} \centering
\resizebox{\hsize}{!}{\includegraphics{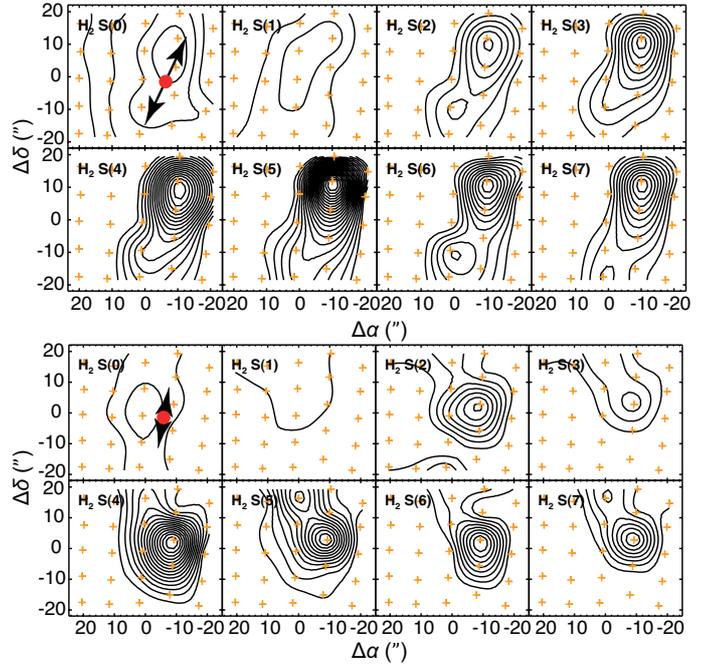}}
\caption{Spitzer spectral maps of the H$_2$ rotational transitions around SMM3
(upper panels) and SMM4 (lower panels), resampled on the PACS footprint pattern.
The 0-0 S(0) and S(1) transitions show diffuse emission. Higher energy
transitions display the outflow pattern as in the case of CO and H$_2$O, except
for SMM4, where the extension to the north is only traced in [OI] (see
Fig.~\ref{fig:9}). Contours start at 10$^{-14}$ erg cm$^{-2}$ s$^{-1}$ and
increase at 10$^{-14}$ erg cm$^{-2}$ s$^{-1}$ steps.} \label{fig:12} \end{figure}

Maps of the H$_2$ rotational ladder from the S(0) to the S(7)
rotational transitions around SMM3 are shown in  the upper panel of
Fig.~\ref{fig:12} after resampling to match the PACS resolution. 
The S(0) and S(1) transitions display emission in a diffuse,
elongated structure, roughly resembling the outflow pattern traced with CO and
H$_2$O and [OI]. Similar diffuse H$_2$ emission has been observed in
protostellar environments for a number of other Spitzer spectral scan studies
\citep[e.g.][]{Maret:09a, Dionatos:10b}. In SMM3, higher energy transitions
between S(2) and S(7) show an outflow distribution similar to that observed in
CO, with a secondary peak at the southern lobe as seen in H$_2$O.

\begin{table*}[!ht] \caption{\textit{Spitzer} - IRS line fluxes measured at
selected positions of peak emission. Reported errors are at 3-sigma level.}
\label{tab:2} \centering \begin{tabular}{l c c c c c c } \hline\hline Element  &
Transition & Wavelength ($\mu$m)   &      \multicolumn{4}{c}{Flux (10 $^{-14}$
erg cm$^{-2}$ s$^{-1}$)} \\ &   &      &   SMM3  b &  SMM3 c   &     SMM3 r  &
SMM4 b \\

\hline \hline

 H$_2$  & $0 - 0$   S(7)    &                               5.5111   &
23.17$\pm$0.83  & 2.94$\pm$1.25 &  5.46$\pm$1.03 &   17.35$\pm$0.77   \\ H$_2$ &
$0 - 0$   S(6)    &                               6.1085  & 11.59$\pm$1.04  &
2.29$\pm$1.45 &  5.34$\pm$1.45 &   9.82$\pm$0.62  \\ H$_2$  & $0 - 0$   S(5) &
6.9095   & 48.55$\pm$1.24  & 5.94$\pm$1.83 & 9.86$\pm$1.90 &   26.30$\pm$1.07
\\ H$_2$  & $0 - 0$   S(4)    & 8.0251   & 16.40$\pm$4.34  & 2.05$\pm$1.10 &
5.42$\pm$1.24 &   17.40$\pm$2.13 \\ H$_2$  & $0 - 0$   S(3)    &
9.6649   & 20.47$\pm$0.43  & 5.74$\pm$0.27 &  6.78$\pm$0.28 &   7.57$\pm$0.47
\\ H$_2$  & $0 - 0$   S(2)    &                               12.278   &
6.58$\pm$0.42  & 1.36$\pm$0.31 &  3.44$\pm$0.49 &   7.75$\pm$0.37   \\ H$_2$  &
$0 - 0$   S(1) &                               17.039   & 5.68$\pm$0.40  &
5.22$\pm$0.59 & 4.51$\pm$0.14 &   2.31$\pm$0.25  \\ $[$SI$]$  & $^{3}$P$_{1}  -
^{3}$P$_{2}$ &    25.249   & 3.90$\pm$0.52  & 4.21$\pm$0.30 &  3.51$\pm$0.24 &
3.51$\pm$0.34 \\ $[$FeII$]$ & $^{6}$D$_{7/2} - ^{6}$D$_{9/2}$        &    25.988
& 2.20$\pm$0.25  & 0.68$\pm$0.18 &  $\ldots$  &   1.83$\pm$0.08    \\ H$_2$ & $0
- 0$   S(0)                                &    28.218   & 3.77$\pm$0.16  &
  2.09$\pm$0.21 &  2.90$\pm$0.17 &   1.52$\pm$0.06  \\ $[$SiII$]$ &
$^{2}$P$_{3/2} - ^{2}$P$_{1/2}$        &    34.815   & 5.17$\pm$0.61  &
2.56$\pm$0.21 &  3.19$\pm$0.22 &   6.24$\pm$1.37  \\

\hline \end{tabular} \end{table*}

Similarly, the resampled IRS map at the PACS scale around SMM4 is shown in the
lower panel of Fig.~\ref{fig:12}.  As in the case of SMM3, the S(0) and S(1)
H$_2$ transitions show diffuse, unconfined emission. The higher energy
transitions  (S(2)-S(7)) present a strong peak located at the same position as
in the PACS molecular line maps and an elongated structure towards the N, having
no clear association with the outflow structure mapped with any other tracer
(e.g. low-J CO, interferometric maps). This is similar to the structure observed
in the $^3$P$_1$-$^3$P$_2$ [OI] transition map (63 $\micron$) and may be
interpreted as gas excited by some of the surrounding protostellar sources.
Besides this exception, the H$_2$ emission around SMM3 and SMM4 shows a pattern 
that is very
similar to those of CO, H$_2$O and [OI]. H$_2$ line fluxes for the same
positions as in Table~\ref{tab:1} are listed in Table~\ref{tab:2}.

\subsubsection{Atomic emission} \label{sec3.2.2}

\begin{figure} \centering
\resizebox{\hsize}{!}{\includegraphics{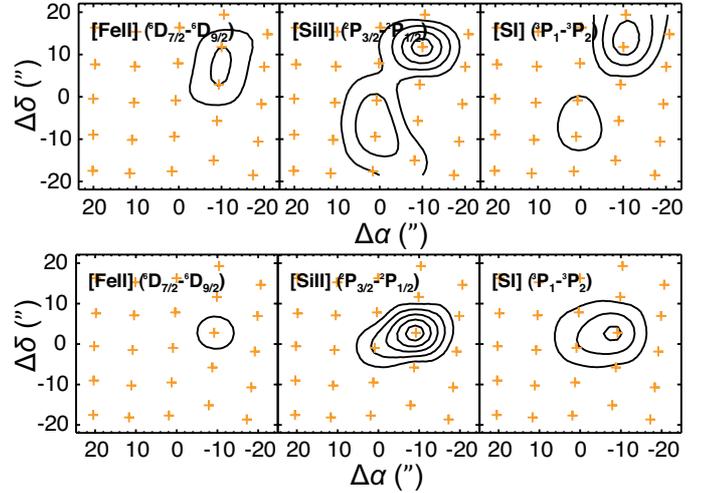}}
\caption{As in Fig.~\ref{fig:12}, for the atomic lines detected by IRS. }
\label{fig:13} \end{figure}

Figure \ref{fig:13} presents line maps from the fundamental transitions of
[FeII], [SiII] and [SI] detected with Spitzer around SMM3 (top panel) and SMM4
(lower panel). [FeII] and [SiII] lines peak at NW of SMM3, at the same region as
most of the molecular and atomic lines. In contrast, the [SI] map around SMM3
delineates the outflow morphology showing clearly both the blue and the
red-shifted lobes. In the case of SMM4, all atomic lines peak far from the
source position, showing a similar morphology as the molecular species traced by
PACS.  Atomic line fluxes are reported in Table~\ref{tab:2}.

\section{Analysis} \label{sec:4}

\subsection{Excitation conditions} \label{sec:4.1}

The physical conditions of the excited gas may be directly constrained by means
of excitation diagrams, i.e., plots of the logarithm of the column density over
the statistical weight as a function of the upper level energy for a range of
transitions of one species. 
%If all the considered lines are optically thin and
%arise in Local Thermodynamic Equilibrium (LTE) conditions in an isothermal
%medium, the distribution of data points on an excitation diagram will be linear.
%[NJE: Ok here is my attempt at clarifying this]
Assuming that the transitions are optically thin, the populations of the
upper states can be calculated simply from the line fluxes and their
logarithms can be plotted versus the upper state energy. If a straight line
can be fitted to these points, a rotational temperature ($T_{rot}$) can
be found from the slope  \citep[e.g.][]{Goldsmith:99a}. 
\trot\ represents the common excitation temperature
of the levels considered and need not be the same as the kinetic temperature
(\tk) \citep[e.g.,][]{Neufeld:12a}, but is the correct temperature to use
in the partition function. Together with the intercept, the partition function
can be used to get the column density or total number of molecules in all
levels.
%In that case, the temperature and column density can be directly calculated from
%the slope and the intercept of the linear distribution with the column density
%axis, given that the partition function of the molecule is known
One can also  estimate the
ortho-to-para ratio of molecules such as H$_2$ and H$_2$O for which a number of
lines are observed in the same wavelength range for both the ortho and para
spin-states.

For the applications discussed in the following, upper level energies and
Einstein coefficients were adopted from the
JPL\footnote{http://spec.jpl.nasa.gov/}  and
CDMS\footnote{http://www.astro.uni-koeln.de/cdms/}
catalogs
\citep[][respectively]{Pickett:98a, Muller:05a}. 
%[NJE: did you get them for the exact Trot or did you have to interpolate?
%I don't really like getting things from tables. See equations in Joel's papers.]
Partition functions for
different temperatures were retrieved from the same databases interpolating for the corresponding \trot\, with the
exception of hydrogen where it was calculated by summing up the first 35 energy
levels of the molecule.

\begin{figure*}[!ht] \centering
\resizebox{\hsize}{!}{\includegraphics{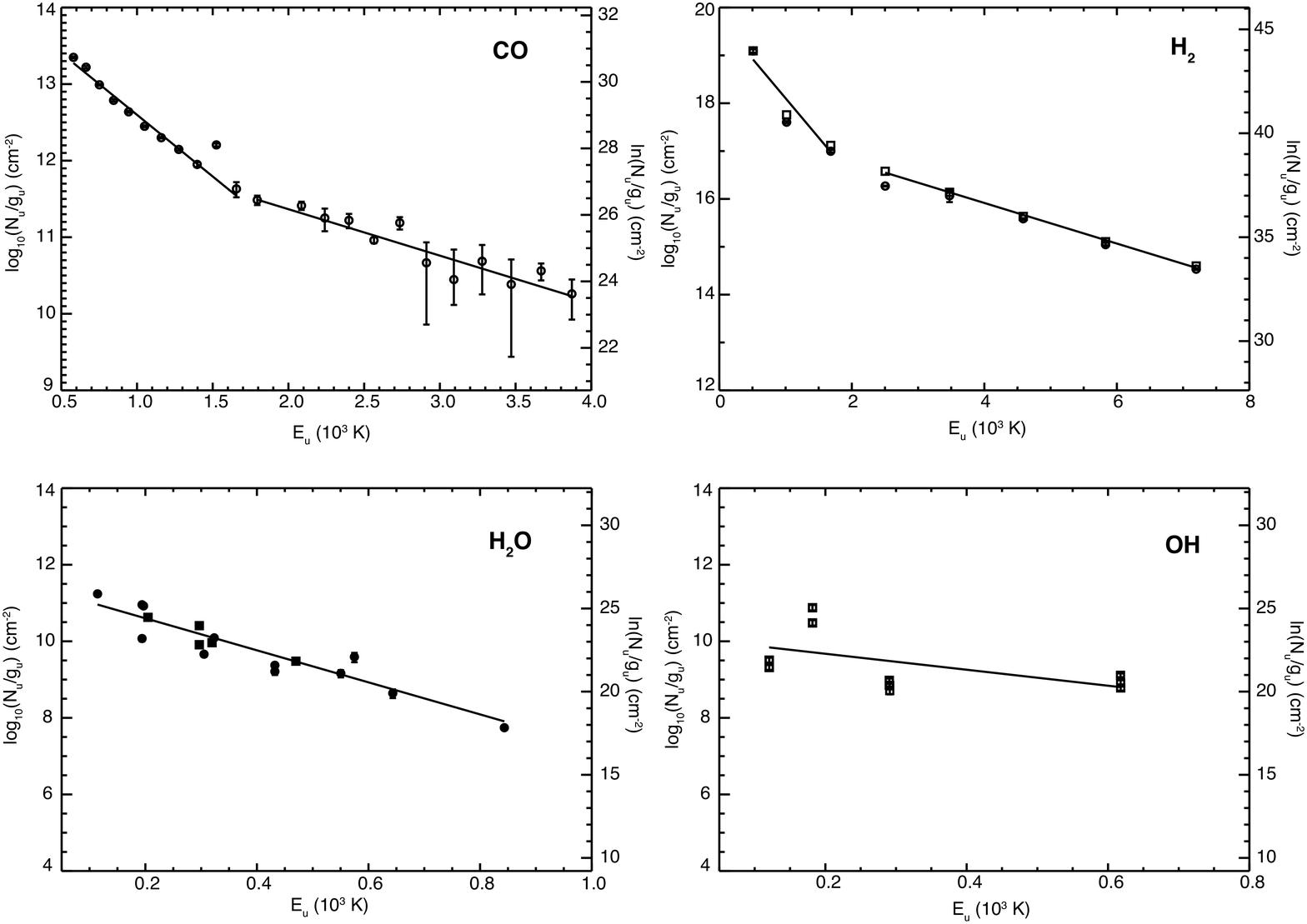}}
\caption{Excitation diagrams for the molecular emission towards SMM3b. Column
densities are given in base 10 and natural logarithms in the left and right hand
sides of each plot. CO (top left) and H$_2$ (top right) plots show a break at
E$_u \sim$1600 K defining a ``warm'' and a ``hot'' gas component at temperatures
$\sim$300K and 900K, respectively. H$_2$O (lower left) and OH (lower right)
display lower temperatures at $\sim$100 K and high scatter of data points, which
indicates that some transitions are optically thick and/or subthermally
populated.} \label{fig:15} \end{figure*}

We employ excitation diagram diagnostics for the CO, H$_2$O and OH lines
detected with Herschel/PACS, as well as H$_2$ detected with Spitzer/IRS to
constrain the physical conditions for each molecule. Column densities were
estimated from the PACS line fluxes reported in Tables~\ref{tab:1} and
\ref{tab:2} focusing on the regions where the molecular emission is most
prominent (see Sect.~\ref{sec:4.0}). Fig.~\ref{fig:15} shows the excitation
diagrams for the blue shifted lobe at NW of SMM3 (SMM3b); diagrams corresponding
to other positions discussed here are presented in Appendix \ref{app:c}.

In the CO excitation diagram (upper left panel of Fig.~\ref{fig:15}) the
distribution of observed data points displays a break for upper level energies
of $E_u\sim$1700 K. This break suggests the existence of two distinct regions
which correspond to different physical conditions. These regions are fit
separately with straight lines, excluding any blended CO lines as listed in
Table~\ref{tab:1}. The two components correspond to temperatures of $\sim$270~K
and $\sim $715~K, and column densities of 1.6 $\times$ 10$^{16}$  and 9.8
$\times$ 10$^{14} $cm$^{-2}$, assuming that the emission fills the spaxel. The
same two-component trend is found for all the other positions examined, with the
break point always located at E$_u\sim$1700 K. Derived temperature and column
density values for the on-source and outflow positions are reported in
Table~\ref{tab:3}, displaying in all cases both a warm ($T \sim 250$ K) and a
hot ($T \sim 800$K) components, with an average of 25 times higher column
density in the warm component.

Fitting the CO data with two linear segments separated at $E_u\sim$1700
K is a commonly used method \citep[e.g.][]{Herczeg:12a, Green:12a} under the
assumption that 
%the gas is thermalized and 
the emission is optically thin.
\citet{Neufeld:12a} has demonstrated that at sufficiently high temperatures and
for densities below the critical, the excitation pattern of CO at a
single kinetic temperature shows a broken power-law pattern that can reproduce the observed distribution.
%follows a power-law distribution that can imitate the observed configuration. 
This possibility has been examined for SMM1 in Serpens
and B335 that display a similar excitation pattern to the Serpens sources of
this work \citep[][respectively]{Goicoechea:12a, Green:12a}. In both cases they
find that  solutions can be obtained for n(H$_2$) = 10$^{4}$~ cm$^{-3}$ and T
$\sim$~3000K. While low densities are predicted by shock models (Sect.~\ref{sec:4.2}),
 high temperatures are possible if CO does not trace the same gas as H$_2$ (see discussion
  below). 
%[NJE: this is a little too glib. You would need to have hard evidence that
%much higher densities are needed, or that these Tk are unrealistic.]
  \citet{Neufeld:12a} also examines the
possibility that a positive curvature can be also obtained for an admixture of
gas at different temperatures. Such conditions are commonly found to occur in
shocks (see Sect.~\ref{sec:4.2}) and are also consistent with the H$_2$
ortho-to-para ratios lying below the equilibrium value of 3, as discussed
below. From the CO excitation diagrams alone it is not possible to distinguish between these solutions.

\begin{table}[!h] \caption{CO rotational temperatures and column densities}
\label{tab:3} \centering \begin{tabular}{l c c c c} \hline\hline Position &
$T_{1}$ (K) & $T_{2}$ (K) & $N_1$ (10$^{16}$ cm$^{-2}$)   &  $N_2$ (10$^{14}$
cm$^{-2}$) \\

\hline \hline

SMM3b  &	  	270$\pm$10 & 715$\pm$75       & 1.6$\pm$0.2 &
9.8$\pm$1.4\\ SMM3c  &		260$\pm$10 & 750$\pm$80     & 0.9$\pm$0.15 &
3.8$\pm$0.9\\ SMM3r   &		230$\pm$10 & 1020$\pm$280       & 1.4$\pm$0.14 &
2.3$\pm$2.5\\ SMM4  &		270$\pm$10 & 720$\pm$50  & 1.8$\pm$0.2 &
9.4$\pm$2.0\\

 \hline \end{tabular} \end{table}

In the case of molecular hydrogen, an estimate of the visual extinction and the
ortho-to-para ratio can be derived following the method described in
\citet{Wilgenbus:00a}.  The statistical weight $g_u$ for H$_2$ includes
the spin degeneracy term ($2S+1$, with $S=0$ for para and $S=1$ for ortho
states). Differences from the LTE value in the ortho-to-para ratio will then be
reflected as vertical displacements between the ortho and para transitions in
the excitation diagram, forming a ``saw-tooth'' pattern. The observed
ortho-to-para ratio is estimated from the displacement of the S(5) line, in
comparison to the levels of S(4) and S(6). For the extinction, the calculation
is based on the fact that  the S(3) pure rotational level at 9.7$\micron$ is
particularly sensitive to the interstellar extinction, as it is located within
the wide-band silicate absorption feature centered at the same wavelength;
therefore column density displacements of this line compared to other hydrogen
transitions can provide information on the extinction, which can be related to
visual extinction assuming an $A_{9.7}$/$A_V$ ratio equal to 0.087
\citep{Rieke:85a}.  
%[NJE: warning. there are issues with this as the silicate feature is now
%known not to trace extinction very well in dense regions.]

The upper right hand panel of Fig.~\ref{fig:15} shows the H$_2$ excitation
diagram at SMM3b, in which observed line fluxes were corrected for an
interstellar extinction of A$_V$=8.2 mag (open squares), derived from the
displacement of the S(3) pure rotational level. In addition, the H$_2$ S(0) and
S(1)  line intensities were corrected for diffuse emission, after estimating an
average value from the positions not associated with known outflows (see Fig.
\ref{fig:a2}).  As in the case of CO, a similar break is seen at
$E_{up}\sim$1500 K, which separates data points into two temperature components,
corresponding to $\sim$260 K  and $\sim$1000 K with estimated column densities
of 6.3 $\times$ 10$^{20}$ and 1.1 $\times$ 10$^{19}$ cm$^{-2}$, respectively.
Excitation diagrams for other positions of interest are presented in Appendix B,
and derived values for the corresponding physical conditions are listed in
Table~\ref{tab:4}. In all cases, molecular hydrogen traces gas separated into a
``warm'' and a ``hot'' component, each one tracing, within uncertainties,
similar column densities between different positions. 
In contrast, SMM4 has an
%there is a clear distinction in the visual extinction and ortho-to-para 
%ratios between the two sources, 
A$_V$ twice as high and ortho-to-para ratio about half of the
equilibrium value of 3 compared to SMM3 (see also Fig.
\ref{fig:b2}). The ortho-to-para ratio is here estimated for lines
tracing the ``hot'' component, and may not be representative for the ``warm''
gas.

The positive curvature observed in the CO excitation diagram may be attributed
in effects other than a distribution of temperatures. As mentioned already,
\citet{Neufeld:12a} has shown that for CO, an isothermal medium with low 
gas density can explain the observed distributions. However,
%In contrast for H$_2$ that shows similar excitation distribution, 
\citet{Neufeld:12a} argues that a similar distribution in H$_2$ cannot be
produced by an isothermal medium. In the next section
we examine the relation of CO to H$_2$, and argue that both molecules trace the
same gas corresponding to a distribution of temperatures rather than a
very hot isothermal medium with low density.

In an analogous study of the outflow emerging from SMM1 in Serpens,
\citet{Goicoechea:12a} find a similar temperature structure, with an additional
cold component traced by CO lines with $J_{up} \leq$ 14 (observed with SPIRE).
Reported column densities are about 2 orders of magnitude higher for both the
``warm'' and ``hot'' components in SMM1 than corresponding ones reported here. 
%[NJE: please clarify whether the 2 orders of mag. higher refers to SMM1 vs
%SMM 3 and 4 or to the cold versus warm and hot components of SMM1]
 This difference can be interpreted in terms of
the smaller emitting area of 4$\arcsec$ assumed and the estimating method, which
is based on non-LTE analysis. The CO column density ratio of the ``hot'' and
``warm'' components in both studies is $\sim$20 indicating that despite the
different approaches, the relative contribution of the two components remains
invariable.  Analysis of Spitzer data around other embedded protostellar
sources \citep[e.g. NGC1333, HH211 and L1448,][respectively]{Maret:09a,
Dionatos:10a, Nisini:10a} find a very similar excitation structure for molecular
hydrogen, indicating that such excitation conditions are common around embedded
protostars.

\begin{table*} \caption{Physical conditions derived from the H$_2$ excitation
diagrams} \label{tab:4} \centering \begin{tabular}{l c c c c c c} \hline\hline
Position &   T$_{1}$ (K) & T$_{2}$ (K) & N$_1$ (10$^{20} cm^{-2}$)   &  N$_2$
(10$^{18} cm^{-2}$) & A$_V$ (mag) & OPR \\

\hline \hline

SMM3b  &	  	260$\pm$8 & 1025$\pm$25 & 8.1$\pm$0.6 & 11.0$\pm$0.7&
8.8 & 3.0\\ SMM3c  &		210$\pm$5 & 970$\pm$50 & 7.5$\pm$0.2 &
1.5$\pm$0.5&  2.0 & 2.8\\ SMM3r   &		250$\pm$5 & 970$\pm$90 &
8.0$\pm$1.3 & 4.7$\pm$1.7& 10.5 & 1.7 \\ SMM4b  &		355$\pm$15 &
1000$\pm$80 & 9.5$\pm$0.4 & 14.0$\pm$4.7 & 20.0 & 1.7 \\

 \hline \end{tabular} \end{table*}

H$_2$O and OH excitation diagrams are presented in the two lower panels of
figure \ref{fig:15}. Both molecules show significant scatter over column
densities. For water this effect may indicate that a number of lines are
optically thick or subthermally populated \citep[e.g.][]{Herczeg:12a}. In the
case of OH, subthermal excitation and IR pumping have been found to be
responsible for the observed scatter \citep{Wampfler:12a}.   Emission from water
vapor around SMM3 and SMM4 is associated with gas at temperatures $\sim$100 K
and column densities of $\sim5-10\times10^{12}$ cm$^{-2}$. Corresponding
temperatures for OH are 50-100\% higher and column densities are about an order
of magnitude lower (see Table~\ref{tab:6}). However, the large scatter in the
diagram unavoidably leads to higher uncertainties, and the values given here
should only be considered as rough estimates.

\begin{table} \caption{H$_2$O and OH rotational temperatures and column
densities.} \label{tab:5} \centering \resizebox{\hsize}{!}{ \begin{tabular}{l c
c c c} \hline\hline Position &   $T$\subscr{H$_2$O} (K) & $T$\subscr{OH} (K) &
$N$\subscr{H$_2$O} (10$^{12}$ cm$^{-2}$) &  $N$\subscr{OH} (10$^{12}$ cm$^{-2}$)
\\

\hline \hline

SMM3b  &	  	105$\pm$20 & 200$\pm$80 & 9.7$\pm$2.2 & 1.1$\pm$1.8\\
SMM3c  &		120$\pm$25 & $\cdots$ & 4.1$\pm$2.8 & $\cdots$\\ SMM3r &
105$\pm$20 & $\cdots$ & 7.6$\pm$4.1 & $\cdots$\\ SMM4b  & 88$\pm$10 & 160$\pm$60
& 8.2$\pm$3.9 & 1.4$\pm$2.3\\

 \hline \end{tabular} } \end{table}

\subsubsection{Relation between CO and H$_2$} \label{sec:4.1.1}

CO is the most common tracer of protostellar outflows. It is easily excited even
at low temperatures found in quiescent molecular clouds ($T\sim$20 K) and its
lower energy transitions (J$_{up}<$  7) are readily accessible from ground-based
observatories. Due to these properties, CO is commonly used as a proxy for the
more abundant H$_2$, which, as a light homonuclear molecule, is only excited in
energetic environments. The determination of the total amount of H$_2$ from
measurements of CO is commonly assessed under the assumption that the CO/H$_2$
abundance ratio is almost constant and equal to 10$^{-4}$, as measured by
\citet{Watson:85a}.

%[NJE: Not clear why this reference is separate from all the other
%references to previous measurements. As I suggested, I would put
%all the discussion of other measures in the intro para. and then focus
%on those most relevant to the hot, shocked gas considered here.]

The majority of available direct measurements of the CO/H$_2$ ratio rely on
observations of far-ultraviolet absorption lines from both molecules
superimposed on the spectra of background stars in diffuse clouds
\citep[e.g.][]{van-Dishoeck:87a}. For dense clouds, attempts to measure the
CO/H$_2$ ratio focus on the simultaneous study of the near-infrared (NIR) CO and
H$_2$ ro-vibrational lines in absorption \citep[e.g.,][]{Lacy:94a} so it is not
obvious that these tracers sample the same volume of gas \citep[for a review of
the topic see also][]{van-Dishoeck:92a}. 

\begin{figure} \centering
\resizebox{\hsize}{!}{\includegraphics{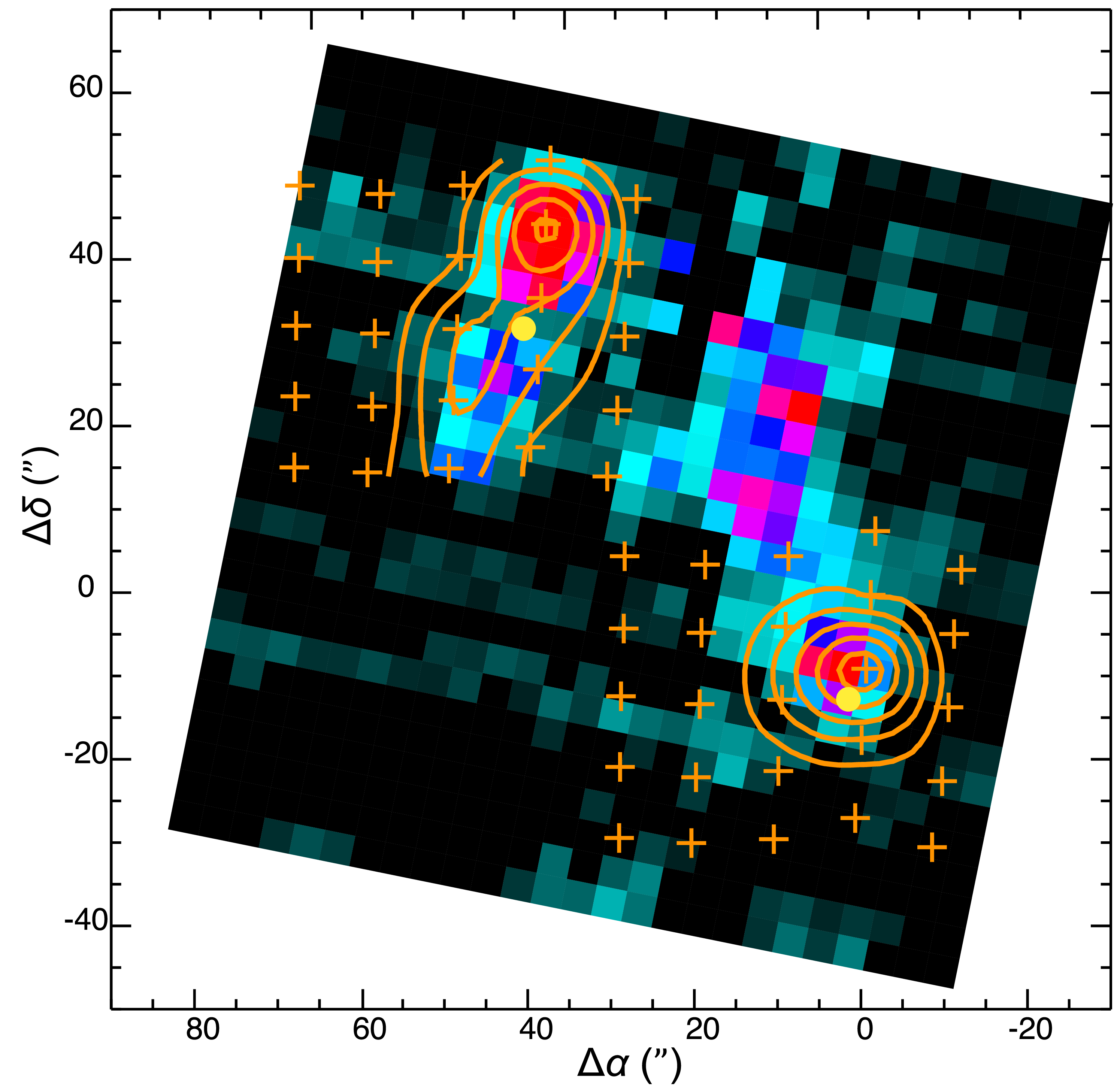}}
\caption{CO $J=18 -17$ emission (orange contours) superimposed on the H$_2$ S(5)
line map around SMM3 and SMM4, displaying the strong spatial correlation between
the two molecular tracers. The positions of SMM3 and SMM4 are marked with filled
yellow circles. Color encoding for the hydrogen map is as in Fig.~\ref{fig:11}
and levels for the carbon monoxide as in Fig.~\ref{fig:5} and \ref{fig:6}. }
\label{fig:16} \end{figure}

The analysis of the CO and H$_2$ excitation conditions (\S \ref{sec:4.1} above)
indicates that their emission originates in gas with very similar
characteristics; both molecules trace a two-temperature structure with a
``warm'' and a ``hot'' component at $T$\subscr{warm}$\sim300$ K and
$T$\subscr{hot}$\sim$1000 K, respectively. At the angular resolution of the
observations (9.4$\arcsec$, PACS) the emission pattern of both molecules is
tightly correlated (Fig.~\ref{fig:16}). This set of common characteristics
suggests that the CO and H$_2$ emission originates from the same physical
processes occurring within the same volume of gas. Therefore the ratio of the
column densities calculated from the excitation analysis corresponds to the
abundance ratio of the two molecules. Still, even the warm component examined
here lies at much higher temperatures than the quiescent gas in molecular clouds
($T\sim20$ K) and therefore the ratios estimated here apply only in highly
excited gas.

\begin{table} \caption{CO/H$_2$ column density ratio} \label{tab:6} \centering
\begin{tabular}{l c c} \hline\hline Position &
($N$\subscr{CO}/$N$\subscr{H$_2$})\subscr{warm} (10$^{-5}$) &
($N$\subscr{CO}/$N$\subscr{H$_2$})\subscr{hot} (10$^{-5}$)\\ \hline

SMM3b  &	  	2.0$\pm$0.16   & 8.9$\pm$1.4  \\ SMM3c  & 1.3$\pm$0.16
& 24.9$\pm$10.3 \\ SMM3r   &		1.8$\pm$0.19   & 4.9$\pm$3.6 \\ SMM4b  &
1.9$\pm$0.15   & 6.7$\pm$2.7  \\

 \hline \end{tabular} \end{table}

Estimates of the CO/H$_2$ ratio are given in Table~\ref{tab:6}. Values around
$\sim$1.8$\times$10$^{-5}$ for the ``warm'' gas and from
$\sim$5$\times$10$^{-5}$ up to $\sim$2.5$\times$10$^{-4}$ for the ``hot'' gas
show a significant discrepancy of the ratio between the two physical components.
Despite the rather high uncertainties, these estimates are indicative of a
higher abundance ratio  by a factor of 3 -- 10 in the ``hot'' gas compared to
the ``warm'' component.

%[NJE: here is my proposed restatement of this without LTE.]
Our estimations rely on the assumptions 
that the lines are optically thin and
%of LTE conditions and
%the observational indications 
that the two molecules trace the same volume of gas. 
The similarity of the temperatures of the two components for CO and H$_2$
support the second assumption. Further, since the H$_2$ levels are likely
in LTE, and the \trot\ of the CO levels are similar for warm
and hot components, the CO is likely in LTE as well ($\tk = \trot$),
making the Neufeld solution unlikely.
%Owing to its low dipole moment, the low-lying levels of the H$_2$ molecule
%are easily thermalized and the LTE assumption may indeed be valid. This however
%does not apply to CO, which has a significantly higher dipole moment and
%therefore its critical density is expected to be much larger, especially for the
%higher frequency transitions. If indeed CO is subthermally excited, then its
%excitation pattern is compatible with low densities and high temperatures which
%are not likely to be found in the cirumstellar gas in the vicinity of an
%embedded protostar (Sect.~\ref{sec:4.1}). Further more our estimations are
%meaningful to the extent the assumption that molecules are excited within the
%same volume of gas is valid.  The derivation of the CO/H$_2$ ratio is subject to
%additional uncertainties. 
For the ``warm'' component, the main source of uncertainty in the CO/H$_2$ ratio
%these are associated more
is the accuracy in the derivation of the H$_2$ column densities, 
while the ``hot''
component is more affected by low signal-to-nose ratio in the CO lines with
$J_{up} > 31$.  
%[NJE: are these incorporated into the uncertainties in Table 6? If so,
%comment. They are quite good for warm, but not so great for hot.]

As discussed in Sect.~\ref{sec:4.1} the line fluxes from the S(0) and
S(1) H$_2$ transitions have been corrected for diffuse emission. This ``cold' '
emission is likely associated with the CO emission arising from $J_{up} < 14$,
tracing larger-scale outflows. After being corrected, the S(0) point may still
maintain some contribution from this colder component (see
Figs.~\ref{fig:15},~\ref{fig:b1}~\ref{fig:b2}). Assuming that this contribution
is still dominating the S(0) emission, we derive physical properties of the
``warm'' component using the S(1) - S(3) lines. In that case the H$_2$
temperature is $\sim 80\%$ higher than reported in Table~\ref{tab:4} ranging
between 400 and 600K. the The corresponding H$_2$ column density is an order of
magnitude lower,  
resulting in a CO/H$_2$ ratio $\sim 2 \times 10^{-4}$, a
factor of 2 higher than ``nominal'' value. For the derived conditions, H$_2$
corresponds to an intermediate component between ``warm'' and ``hot'', which
renders invalid our assumption that CO and H$_2$ trace gas at very similar
conditions. Therefore the ratio obtained by 
discarding the S(0) sets an upper limit to our estimations of CO/H$_2$.
 
%[NJE I could not understand most of thie paragraph.]
The high levels of the S(0) line relative to the S(1) and S(2) can also
be interpreted in terms of non-equilibrium values of the ortho-to-para ratio.
The ratio depends strongly on temperature, and can be significantly smaller than the equilibrium value of 3 for
colder gas \citep[e.g.][]{Neufeld:09a}. 
For non-equilibrium values of the ortho-to-para ratio will be reflected as a vertical displacement between the ortho and para
transitions in the excitation diagram.
In that case, it is the S(1) data point lying at lower levels with respect to the S(0), rather than the latter data point being affected by diffuse emission.
 The ortho-to-para ratio
values estimated here are representative of the ``hot'' gas and may not apply
for the ``warm'' component. If the ``hot'' gas has not reached the equilibrium
ortho-to-para value, it would be even more so for the ``warm'' component.
Indeed, for the positions that the ``hot'' ortho-to-para ratio is found below 3,
the relative displacement between the S(0) and S(1) lines even more pronounced.
(Figs.~\ref{fig:b1},~\ref{fig:b2}). In this case the derived ``warm'' H$_2$
column densities would be underestimated and the CO/H$_2$ ratio in the ``warm''
gas would be lower.

%[NJE As I said, move this to intro and focus on relevant ones]
In comparison with analogous studies our estimates  are different compared to the ratio of
$\sim2\times$10$^{-4}$ found for $\sim$50 K gas in \citet{Lacy:94a}, but fall
within the limits of log(CO/H$_2$) between -7.58 and -4.68 measured by
\citet{Sheffer:08a}.

\subsection{Comparison with shock models} \label{sec:4.2}

The outflow morphologies of the maps in Sect.~\ref{sec:3} suggest that the bulk
of the line emission is likely produced in shocks.  Emission is traced down to
$\sim$10$\arcsec$ from the source, which corresponds to $\sim$4000 A.U. for the
adopted distance. Therefore some contribution from gas  in the envelope which is
heated-up by UV photons from the protostar cannot be excluded, albeit this is
not suggested by the [C~II] morphology. 
%[NJE: but we know that CII is different, so why bring it up here?]
 In order to further examine these
arguments and to constrain the underlying physical conditions, we compare the
observed line emission to the shock model predictions of \citet{Kaufman:96a} and
\citet{Flower:10a} (hereafter KN'96 and FPDF'10, respectively). Both models
provide line intensity predictions for H$_2$, CO, and H$_2$O for steady-state
shocks. KN'96 models include a more extensive and refined grid of shock
velocities and pre-shock densities but only for C-type shocks, while FPDF'10
models include predictions for J-type shocks, however for a coarser grid of
input parameters. Major differences between the two codes is in the treatment of
the chemistry and the collisional rate coefficients for the CO and H$_2$O.  As
reported in FPDF'10, their models include charged grains, which act to compress
the shock front and increase the temperatures reached. The basic set of values
for the grids of predictions provided by the two models are summarized in
Table~\ref{tab:7a}. For the purposes of the current study, the predictions
provided by the codes can be treated as complementary.

\begin{table} \caption{Parameters of the shock models employed} \label{tab:7a}
{\centering \begin{tabular}{l c c } \hline\hline Model & KN'96 & FPDF'10 \\
\hline v$_s$ (range, km s$^{-1}$)  & $5 - 40$ & $10 - 40^{a}$\\ v$_s$ (step, km
s$^{-1}$) & 5 & 10\\ n (range, cm$^{-3}$) & 10$^4 - 10^{6.5}$ & $2 \times 10^4 -
2 \times 10^5$ \\ n (step, cm$^{-3}$) & 10$^{0.5}$ & 10 \\ J-shocks & no & yes\\
H$_2$ lines & full & full\\ CO lines & full & J$_{up}\leq$20 \\ H$_2$O lines &
limited & full \\ $[$OI$]$ lines & no & full\\ \hline \end{tabular} } \\ {}~$^a$
$10 - 30$ km s$^{-1}$ for J-shocks\\ \end{table}

Model grids are compared to line intensities through $\chi^{2}$  fits to
optimally reproduce the observed line emission for different molecules.
Comparisons are performed considering relative line intensities, introducing a
scaling factor (beam filling factor). This factor accounts for the possibility
 that
the emitting region is smaller than the instrumental beam, which equals to the
PACS spaxel size in this study.  As a general trend, the C-shock models can
reproduce the observed line intensities only if model predictions are scaled for
a small beam-filling factor, while J-shock models require a scaling factor much
closer to unity. The beam filling factor along with the uncertainties in the
fluxes of the observed lines result in degenerate model solutions.

\subsubsection{H$_2$ \& CO}

\begin{figure*} \centering
\resizebox{\hsize}{!}{\includegraphics{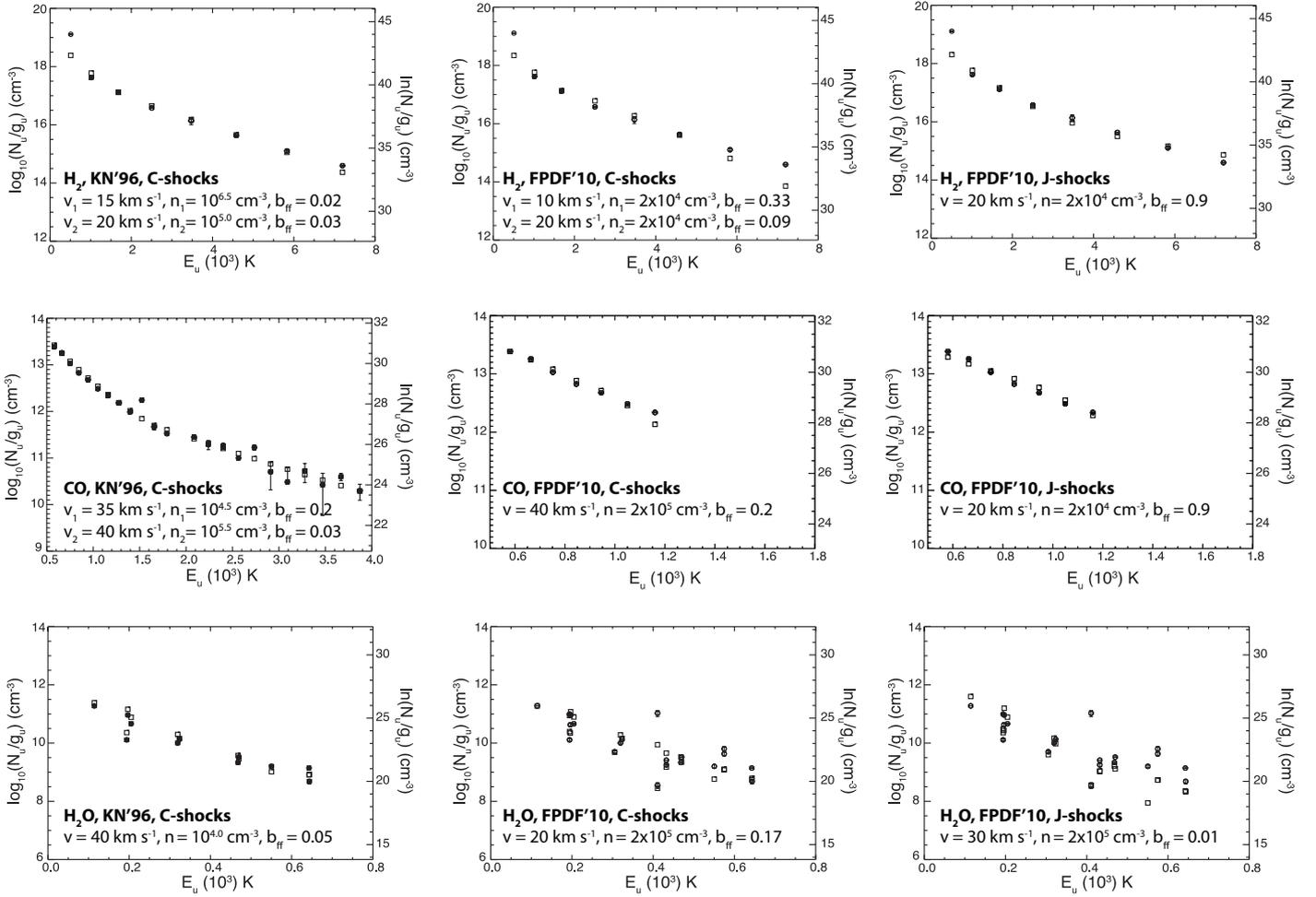}}
\caption{Comparison between observed and best-fitting model intensities for
H$_2$, CO and H$_2$O (top, middle and lower panels, respectively) displayed in
excitation diagrams. C-type shock models  from  KN'96  and FPDF'10 are presented
in the top and center columns, and J-type shocks from FPDF'10 in the right
column. Best fitting model parameters are reported on each panel.}
\label{fig:17} \end{figure*}

Comparisons between H$_2$ and CO intensities at SMM3b and best fitting
models are presented in Fig.~ \ref{fig:17} (upper and middle panels).  As a
general trend we find that C-shock models cannot simultaneously reproduce the
full range of observed lines. The only exception is the FPDF'10 CO model
predictions, which are limited to $J_{up}=20$ and therefore do not cover to
the same extent the CO transitions observed. For C-shocks, the transitions
corresponding to the two temperature components are therefore fitted
independently. For each molecule and temperature component, best fitting models
provide similar solutions for the density and shock velocity at each position of
peak intensity, reflecting a consistent curvature and slope in the excitation
diagram between the different regions. In contrast, the beam filling factor
which scales the model intensities (as expressed in terms of column densities
over degeneracy in the excitation diagrams) relative to the observed ones varies
up to two orders of magnitude between different positions and temperature
components (see Tab.~\ref{tab:8}).

C-shock models predictions from KN'96 and FPDF'10  provide solutions
which are substantially different for the two molecules, or even the two
temperature components of the same molecule. Degenerate model solutions
complicate significantly the interpretation of the predicted values, however a
few trends can be recognized.  The two models provide consistent predictions on
the shock velocity. This ranges from low ($10-15$ km s$^{-1}$) and moderate (20
km s$^{-1}$) values for the ``'warm'' and ``hot''  H$_2$ components, 
to high values
($30-40$ km s$^{-1}$) for all CO temperature components. For both models, beam
filling factors vary up to an order of magnitude even for the same set of
velocities and pre-shock densities in order to accommodate the observed values
at different positions along the outflows. Furthermore, lower beam filling
factor values correspond to higher densities and {\it vice versa} indicating
degenerate solutions. In the case of CO, KN'96 models provide equally good
(degenerate) solutions for a set of models with stepwise increasing density by
0.5 dex and decreasing velocity by 5 km s$^{-1}$.       

 Predictions for J-shock models from the grid of FPDF'10 are presented
on the upper-right and middle panels of Fig.~\ref{fig:17}. Single-temperature
J-shock models are able to reproduce the full range of H$_2$ transitions (with
the exception of S(0)). The corresponding models imply  low pre-shock densities
of 2$\times10^{4}$ cm$^{-3}$ and a moderate shock velocity of 20 km s$^{-1}$,
for all positions examined. For CO, predicted line intensities in the FPDF'10
grid are limited to $J_{up}=20$ so no secure conclusions can be made on the
shock type. However comparisons show that  J-type shocks provide essentially the
same set of solutions for both the CO and H$_2$ emission. Beam filling factors
vary from 30\% to 90\%, with higher values corresponding to the points of peak
emission. Such filling factors can be explained by a chain of unresolved shock
knots within the PACS spaxel size. 

The emission in the 4.5$\micron$ IRAC image presented in
Fig.~\ref{fig:9} (upper left panel) is dominated by higher excitation H$_2$
transitions \citep{Neufeld:06a} and therefore can give a rough measure of the
size of the emitting regions.  The bright emission structures around SMM3b and
SMM4b appear to be close to 10-30\% of the PACS spaxel size and therefore higher
beam filling factors predicted by the FPDF'10 models are more plausible in
describing better the physical conditions therein.  For the positions of SMM3c
and SMM3r, models predict lower beam filling  in comparison to the regions of
brightest emission which is consistent with the smaller/dimmer structures
observed in the IRAC image. In conclusion, J-shock models with pre-shock densities of 2$\times10^{4}$ cm$^{-3}$ and velocities of 20 km s$^{-1}$ can simultaneously reproduce both CO and H$_2$ emission. The predicted size of the emission region from these solutions is consistent with those indicated by imaging.

%%%%%%%%%% New merged table for shock models %%%%%%%%%%%%%

\begin{table*} \caption{Comparison of molecular emission with shock
models.} \label{tab:8} {\centering
%\resizebox{\hsize}{!}{
\begin{tabular}{l c c c c c c } \hline\hline Model & Shock & Molecule &
Component & v$_s$ (km s$^{-1}$)   & n (cm$^{-3}$) & b$_{ff}$ \\ \hline KN'96 & C
& H$_2$ & ``warm'' & 15 & 10$^{6.5}$ & 0.001 - 0.02\\ KN'96 & C & H$_2$ &
``hot'' & 20 & 10$^{5.0} - 10^{5.5}$ & 0.003 - 0.03\\ FPDF'10 & C & H$_2$ &
``warm'' & 10 & $2 \times10^{4.0}$ & 0.14 - 0.35\\ FPDF'10 & C & H$_2$ & ``hot''
& 20 & $2 \times10^{4.0}$ & 0.01 - 0.09\\ FPDF'10 & J & H$_2$ & ``'warm \& hot''
& 20 - 30 & $2 \times10^{4.0}$ & 0.32 - 0.90\\ KN'96 & C & CO & ``warm'' & 30 -
35 & 10$^{4.5}$ & 0.10 - 0.24\\ KN'96 & C & CO & ``hot'' & 20 - 40 & 10$^{5.0} -
10^{6.5}$& 0.001 - 0.03\\ FPDF'10$^{a}$ & C & CO & ``warm'' & 40 & $2
\times10^{5.0}$ & 0.09 - 0.19\\ FPDF'10 & J & CO & ``warm \& hot'' & 20 & $2
\times10^{4.0}$ & 0.50 - 0.98\\ KN'96 & C &H$_2$O & ``cold'' & 10 - 40 &
10$^{4.0} - 10^{5.0}$& 0.05 - 0.21\\ FPDF'10 & C &H$_2$O & ``cold'' & 10 - 20 &
$2 \times10^{4.0} - 2 \times10^{5.0}$& 0.17 - 0.46\\ FPDF'10 & J &H$_2$O &
``cold'' & 30 & $2 \times10^{5.0}$& 0.01 - 0.1\\ 

\hline \end{tabular} \\ } % This is for the centering of the table - do not
{}~$^a$ Comparison limited to $J_{up}=20$ due to available model predictions\\
\end{table*}

%%%%%%%%%%%%%%%%%%%%%%%%%%%%%%%%%%%%%%%%%%

\subsubsection{H$_2$O \& OH}

The lower panels of Fig.~\ref{fig:17} presents the best fitting models from the
KN'96 grid in comparison with the H$_2$O observations. Since the KN'96 model
grid was intended for comparisons to ISO observations, the number of water lines
available is limited to the brightest transitions. Values derived from the
comparison suggest a C-type shock of low density and high velocity (10$^{4}$
cm$^{-3}$, 30 -- 40 km s$^{-1}$, respectively). Degenerate solutions include
models with moderate velocities of 20 km s$^{-1}$ and slightly higher densities
of 10$^{4.5}$ cm$^{-3}$. Best-fitting models are indicative of conditions very
similar to those from the CO comparisons, for the warm gas, however for filling
factors $\sim$10 times lower.

The FPDF'10 C-type grid provides a larger set of lines for comparison (lower
center panel of Fig.~\ref{fig:17}); however even best matching models do not
reproduce sufficiently the observed lines in the high-energy regime.  The
solutions provided are highly degenerate and exclude only the highest density
and velocity provided in the grid (40 km s$^{-1}$ and 2$\times 10^{5}$
cm$^{-3}$).  J-type shock comparisons (lower right panel of Fig.~\ref{fig:17})
do not reproduce the observations, and show a much larger scatter than the
C-type shocks.

The OH/H$_2$O ratio can be used as indicator of the shock type, as shock models
show that it varies from 10$^{-2}$ in C-shocks \citep[e.g.][]{Kaufman:96a} to
10$^{2}$ in the high-velocity J-shocks \citep[e.g.][]{Neufeld:89a}.
The low-velocity J-shock models of Flower et al. do not give predictions
for OH, however it is reasonable to expect that the OH/H$_2$O ratio would be lower
in these type of shocks. Even though the H$_2$O and OH column densities
measured here are rather rough estimates, the large variation of their ratio
predicted in extreme limits of shock conditions allows us to use them
as a possible indicator. Assuming that emission from both molecules
arises within the same volume of gas at the positions of peak emission, the
OH/H$_2$O ratio is $\sim$0.1. This is in favor of a C-type shock but
only to the lowest limit of velocities estimated from the KN'96 models. The best
fit solutions for the majority of the peaks indicate higher shock velocities,
where the OH/H$_2$O ratio is expected to be $<$ 0.01. The OH line flux at 119.23
$\micron$ agrees with the predictions of \citet{Kaufman:96a} for the densities
and velocities derived for H$_2$O in C-shocks.  In contrast, the sum of fluxes
over all OH lines in the PACS range is more than an order of magnitude lower
than the values predicted for J-shocks \citep{Hollenbach:89a}, even if scaled
for the corresponding range of filling factors. In conclusion, shock models and line ratios suggest that the H$_2$O and OH emission is likely to arise in C-type shocks.

\subsubsection{Atomic emission}

Oxygen in shocks is produced in large amounts through sputtering from dust
grains and dissociation of molecules. The efficiency of this process depends
largely on the shock velocity and shock type \citep{Hollenbach:89a}. Models
predict high abundances especially of  the $^3$P$_1$-$^3$P$_2$ [OI] line at
63$\micron$ in the case of dissociative J-type shocks; to test this, a number of
diagnostic line ratios have been proposed \citep{Hollenbach:89a, Flower:10a}.
The shock origin of the 63$\micron$ oxygen line is suggested here from its
spatial distribution and close correlation to the outflows (see Sect.
\ref{sec:3}). Therefore, the [OI] emission provides a unique probe of the
occurrence of J-type shocks.

\begin{figure} \centering
\resizebox{\hsize}{!}{\includegraphics{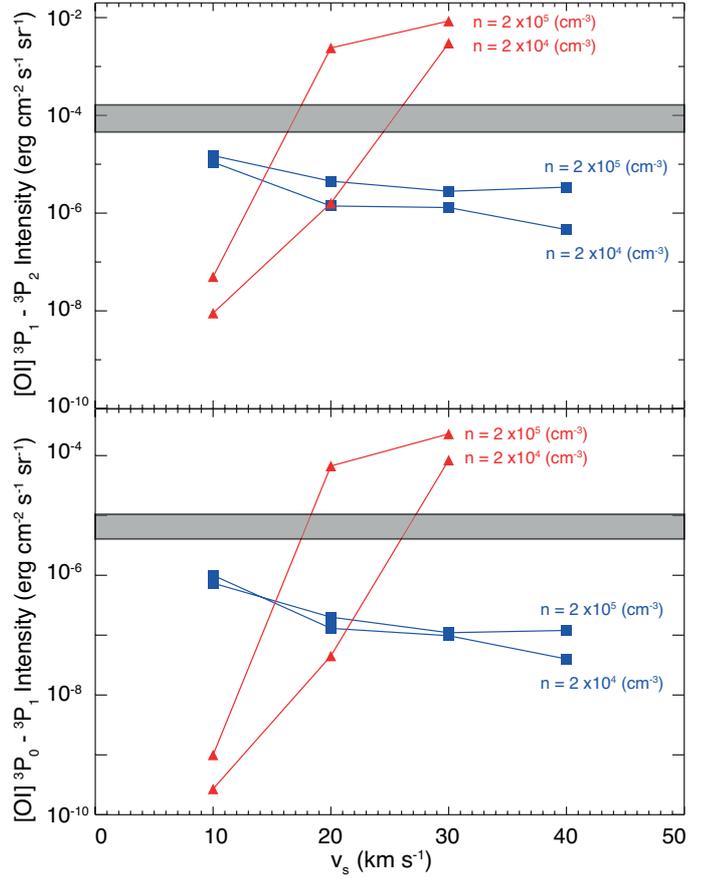}}
\caption{[OI] line intensities for the $^3$P$_1$ - $^3$P$_2$ and the $^3$P$_0$
- $^3$P$_1$ transitions (\textit{upper} and \textit{lower} panels, respectively)
  versus shock velocity, for C- (blue) and J-type shocks (red) from FPDF'10 .
The shaded area indicates the observed upper and lower line intensity limits.
Observed intensities intercept only the J-shock predictions, constraining the
shock velocities to 15-25 km s$^{-1}$. } \label{fig:20} \end{figure}

%[NJE: This comparison is worrisome because most [CII] emission here
%has a different origin. You should address how you separate from the
%PDR emission.]
\citet{Hollenbach:89a} proposed that the [OI](63$\micron$)/[CII](158$\micron$)
intensity ratio would be expected to be below 10 if the emission is attributed
to a photon-dominated region, and greater than 10 for J-type shocks. This is due
partly to the high post-shock densities and temperatures, which favor [OI]
production and partly by the rapid conversion of [CII] to CO in the high
temperature post-shock chemistry which suppresses the [CII] emission. In the
data presented here, the  [OI](63$\micron$)/[CII](158$\micron$) ratio is $>$20
in the brightest regions (SMM3b and SMM4b), and drops to about 7 at SMM3r (see
Table~\ref{tab:1}). This suggests that SMM3r is likely more affected by the diffuse [CII] emission
from SMM6 (see Fig. \ref{fig:9}).

The model grid of \citet{Flower:10a} includes predictions for the [OI] lines at
both 63$\micron$ and 145$\micron$. Figure \ref{fig:20} shows model-predicted
line intensities against shock velocities; for the two densities taken into
account in the model grid, [OI] model intensities define a locus for each
shock-type. On the same figure, shaded areas defined by the minimum and maximum
observed intensities are not corrected for beam filling factors and
therefore represent lower limits in the case the emission is more compact. They
stand well above the C-shock locus and define a common area with the
 J-type shocks, constraining for the given densities the
shock velocity between 15-25 km s$^{-1}$, which agrees well with the velocity
found for the best-fitting J-type shocks in the case of H$_2$ and CO.
Correcting the observed intensities for the emitting region would 
result in higher shock velocities by $\sim$ 5 km s$^{-1}$.

\begin{figure} \centering
\resizebox{\hsize}{!}{\includegraphics{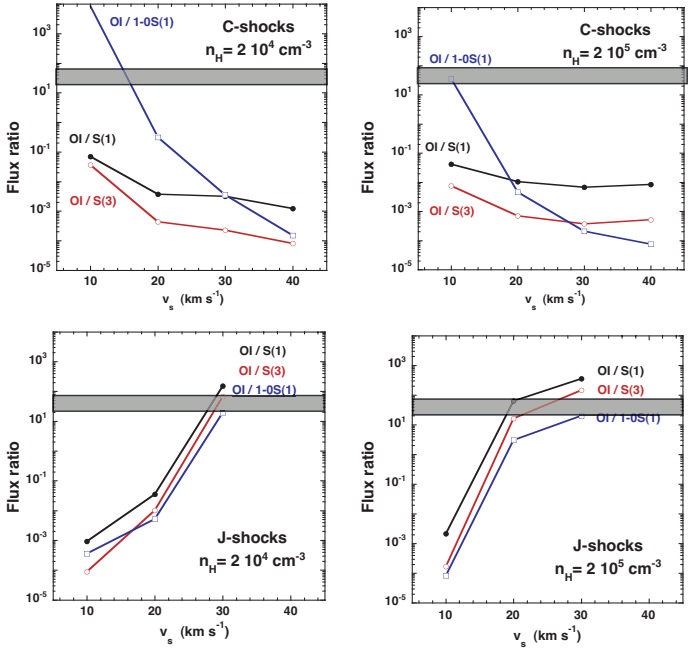}}
\caption{Ratios of the [OI] $^3$P$_1$ - $^3$P$_2$ over the H$_2$ S(1) and S(3)
lines, for pre-shock densities of 2$\times$ 10$^{4}$ and 2$\times$ 10$^{5}$
cm$^{-3}$ (left and right panels, respectively) and C- or J-type shocks (upper
and lower panels). Shaded areas show the observed upper and lower limits of the
ratio for both H$_2$ transitions. The observed ratio overlaps only with J-shock
model predictions, for velocities between 20-25 km s$^{-1}$. (\textit{Figure
adopted from FPDF'10})} \label{fig:21} \end{figure}

As pointed out by \citet{Flower:10a} collisional dissociation of H$_2$ in J-type
shocks leads to the chemical dissociation through reactions with atomic hydrogen
of H$_2$O, OH, O$_2$ and even CO which become rapid only at high kinetic
temperatures reached by J-type shocks. As a diagnostic to the shock type,
\citet{Flower:10a} suggest accordingly the [OI]/H$_2$ ratio which tends to
increase monotonically with the shock velocity for J-type shocks, but to
decrease for C-type ones, assuming that H$_2$ and [OI] are co-existent.  Figure
\ref{fig:21} is a reproduction of Fig. 7 of \citet{Flower:10a} showing the [OI]
over H$_2$ S(1) and S(3) line ratio predictions for a set of C and J shocks with
the corresponding minimum and maximum values for the observed ratios overlaid.
As suggested, the observed line ratio is consistent only with J-type shocks, at
velocities of $\sim$ 20-25 km s$^{-1}$ for densities between 2$\times 10^4$ and
2$\times 10^5$ cm$^{-3}$.

The existence of dissociative shocks is also supported by the detection of [SI]
at 25.2 $\micron$, [FeII] at 26 $\micron$ and  the enhancement  of [SiII] at
34.8 $\micron$  at positions coincident with the [OI] peaks.  Estimates from
\citet{Hollenbach:89a} for J-shocks of $u_s \sim 30$~km~s$^{-1}$ and $n = 10^4$
cm$^{3}$ show that [SI], [FeII], [SiII] and [OI] at 145 $\micron$ lines have
similar intensities between 1 and 4 $\times  10^{-4}$ erg cm$^{-2}$ s$^{-1}$
sr$^{-1}$, while the [OI] line at 63 $\micron$ is more than an order of
magnitude brighter ($\sim7 \times 10^{-3}$ erg~cm$^{-2}$~s$^{-1}$~sr$^{-1}$) and
[CII] is negligible. These results reproduce accurately the measured line fluxes
at the peak positions reported in Tables \ref{tab:1} and \ref{tab:2},   for a
beam ranging between 10 -- 20\% of the PACS spaxel size.

\section{Discussion} \label{sec:5}

Ample evidence for the occurrence of J-type shocks is provided by a number of
different diagnostics.  The high [OI] flux and the [OI]/H$_2$ and [OI]/[CII]
ratios when confronted with models can only be interpreted by J-shocks.  In
addition, the relative flux levels of [FeII], [SiII], [SI], and the two [OI]
lines are consistent with the J-type shock predictions of
\citep{Hollenbach:89a}.  Further evidence stems from the comparisons of the
H$_2$ and CO emission (the latter for $J_{up} \le 20$ ) in a single-shock
scheme. All diagnostics converge to a common set of values with velocities
ranging from 20 to 30 km s$^{-1}$, pre-shock densities between 10$^4$ and 10$^5$
cm$^{-3}$, and beam filling factors of 10 -- 90\%. 
%[NJE: I can't understand what this phrase means.]
To the upper limit, such
filling factors are hard to interpret, however they may occur when
comparing a projected 3-D bow-shock structure on the plane of the sky with 1-D
model data.  

C- type shocks are suggested from the comparisons with the emission from water,
the OH/H$_2$O ratio and the low OH flux levels. C-shocks can reproduce the
emission from CO and H$_2$, however in a piecewise fashion with two shock
components corresponding to the ``warm'' and ``hot'' gas.  The warm gas emission
from CO and H$_2$ along with H$_2$O can be reproduced by models with high
velocities (30 -- 40 km s$^{-1}$) and low-to-intermediate densities (10$^{4}$ -
10$^{5}$ cm$^{-3}$). For the hot gas, predicted shock conditions vary
significantly between different molecules so that it is difficult to define a
common set of solutions.  Beam filling factors vary between 0.1\% and 30\% for
the ``hot'' and ``warm'' gas.  The scatter in the shock conditions can be
interpreted as the emission originating from a projected, two-dimensional
bow-shock. In this scenario, higher excitation lines would originate from a
higher temperature gas closer to the bow-head, where compression is higher and
therefore filling factors are low. In a similar fashion, lower excitation lines
would originate from the bow-shock wings where compression is lower and thus
filling factors higher. The two-dimensional bow-shock scenario can also
accommodate the large span in shock velocities and densities retrieved from
model comparisons for different molecules. J-shocks cannot fit this picture
though, as in contrast to the values derived here beam filling factors would
need to be smaller than C-shocks. Such a scenario could be reconciled
with a non-steady, truncated C-shock with an embedded magnetic precursor (C+J
shocks). However such models were not available for comparisons.

In a recent study, \citet{Benedettini:12a} demonstrated that the [OI], OH and CO
emission traced with PACS in the B1 region on the outflow of L1157 is consistent
only with dissociative J-shocks. It is worth noticing that the [OI] levels in
L1157 are similar to the ones observed here, however CO is only detected up to
$J_{up} = 22$. Therefore at least the atomic lines here and in the study of
\citet{Benedettini:12a} provide strong indications for the existence of J-type
shocks. Another study focusing on HH54 \citep{Bjerkeli:11a}, concludes that the
observed cooling rates between CO and water can only be reconciled if J-shocks
are taken into account. Concerning the molecular lines, a similar comparison of
rotational H$_2$ emission in the protostellar outflow from HH211
\citep{Dionatos:10a} has provided consistent results with the current study. C-
and J-type shocks can account for the observed line intensities considering low
and high beam filling factors, respectively.  Comparisons of CO and H$_2$
emissions from ISO and ground based facilities in the case of HH54
\citep{Giannini:06a} have concluded that C-type shocks are not sufficient to
reproduce the observed intensities from both molecules.  \citet{Giannini:06a}
find that a steady state J-type or a quasi-steady J-type shock with a magnetic
precursor (C+J shock) can best fit the observed line intensities.  More
recently, water emission observed with HIFI \citep{Santangelo:12a} was suggested
to originate in J-shocks, which is not consistent with the findings from the
shock model comparisons above. However, \citet{Kristensen:12a} has shown that
the water line profiles resolved with HIFI are complex, tracing simultaneously
more than one processes.

%Except from shocks, more m
Mechanisms other than shocks may be responsible for the excitation of
gas. \citet{Visser:12a} have demonstrated that CO emission in the PACS range can
be reproduced partly by UV heating of the envelope gas and partly by C- shocks
acting on the outflow cavity walls. Around SMM3 and SMM4, the bulk of [CII]
emission is associated with SMM6, however some [CII] emission is also observed
along the outflows.  A protostellar source with UV luminosity of $\sim$ 0.1
L$_{\odot}$, as indicated by best-fitting model-sources of \citet{Visser:12a}
would give an unattenuated UV flux of $\sim$ 10 $G_0$ at a distance of 4000 AU
from the source (e.g. the distance of the outflow positions from SMM3). Adopting
an envelope density of  $\sim 10^4$ cm$^{-3}$ as found in \citet{Visser:12a},
and using the PDR model of \citet{Kafuman:99a} we find [CII] flux levels of
$\sim$ 10$^{-14}$ erg cm$^{-2}$ s$^{-1}$ within the area of a PACS spaxel. This
estimate corresponds well to the values reported in Table \ref{tab:1} (0.5 --
2.5 $\times 10^{-14}$ erg cm$^{-2}$ s$^{-1}$). 
%[NJE: same worry as above about using CII.]
However, in this scenario where
the envelope gas is heated by UV radiation from the protostar, the [CII] flux
should drop as a function of distance from the exciting source. As reported in
Table \ref{tab:1}, the [CII] flux levels at the outflow peaks are $\sim$4 times
higher than the spaxel closest to the source (SMM3c), Therefore the gas
excitation due to UV radiation from the protostar is not consistent with the
distribution of the [CII] flux demonstrated by the current data. If [CII]
excitation is attributed mostly to SMM6, then lower UV levels from SMM3 would
play a minor role in the gas excitation at the distances examined.

The analysis in Sect. \ref{sec:3} and Sect. \ref{sec:4}  suggests a common
origin for the H$_2$ and CO emission. The observed two-temperature component may
be interpreted by either two component C- or single component J-type shocks. The
difference by a factor between 3 and 10 in CO/H$_2$ ratio measured for the warm
and hot gas components cannot be easily interpreted by a C-shock chemistry, as
such shocks are  non-dissociative and therefore the abundances of major
molecular gas constituents are not expected to change significantly. In
addition, sputtering of ice mantles from the dust grains is a very efficient
process in C-shocks with velocities greater than $\sim$ 15 km s$^{-1}$, due to
the differential speed between the neutral and the charged fluids. Therefore
C-shocks cannot easily interpret the observed variation in the CO/H$_2$ ratio,
either through chemical reactions or depletion through the formation of ices
onto dust grains.

The H$_2$ binding energy is almost 3 times lower than that of CO; therefore
dissociative J-shocks should in principle have a more destructive impact
on H$_2$ rather than CO molecules. Once H$_2$ is dissociated though, CO can be
efficiently  dissociated in chemical reactions with atomic hydrogen
\citep{Hollenbach:89a}. Therefore in a fully dissociative shock, both CO ands
H$_2$ emission would originate in the post-shock gas, when temperatures drop
enough to allow molecules to reform. H$_2$ reformation occurs on dust grains,
and therefore the efficiency of H$_2$ production depends on the survival of dust
in the post-shock gas. On the other hand, CO can be produced rapidly through
fast-neutral reactions in the gas phase, as long as the temperature remains
higher than 300 K \citep[see Fig.2 in ][]{Hollenbach:89a}. As the post-shock gas
cools down CO reforms and is eventually adsorbed onto dust grains, where H$_2$
is produced and ejected into the gas phase \citep{Flower:10a}.  In this scheme,
J-type shocks can explain the CO abundance variation; CO is more abundant in the
hot gas where H$_2$ is still partially dissociated and CO forms through fast
gas-phase reactions.  The warm gas abundance ratio is then consistent with the
CO being adsorbed and the H$_2$ desorbed from/onto dust grains as the gas cools
down. The J-shock scenario seems to provide explanation to the CO abundance
variation, however there are two caveats. First, CO dissociation is expected to
produce strong [CI] emission which is not observed on the outflows in
APEX/CHAMP+ maps of Serperns (L E. Kristensen, in prep.). Furthermore, the
shock conditions and beam filling factors reported in Table \ref{tab:8} for
H$_2$ and CO are essentially identical, which may indicate intrinsic weaknesses
of the comparisons and the model solutions. 

In a different scenario, the hot and warm gas emission may represent cooling due
to J- and C-type shocks, respectivelly. As mentioned in Sect.~\ref{sec:3}, the
temperature break in H$_2$ and CO is observed in a large number of outflows, and
is likely to reflect an omnipresent phenomenon related to the underlying
physical processes. To test this, a larger sample of protostellar outflows and
more detailed shock models are necessary.

\section{Conclusions} \label{sec:6}

We have carried out spectro-imaging observations of SMM3 and SMM4 in Serpens
with Herschel/PACS and Spitzer/IRS. These observations provide an almost
complete wavelength coverage from 5 $\micron$ to 190 $\micron$, at angular
resolutions of $\sim$9.4$\arcsec$ and reveal a wealth of lines originating from
rotational transitions of H$_2$, CO, H$_2$O, OH, as well as forbidden
transitions of [OI], [CII], [FeII], [SI] and [SiII]. The main results are
summarized as follows:

\begin{itemize}

\item{The morphology of molecular (H$_2$, CO, H$_2$O and OH) and atomic emission
([OI], [FeII], [SiII], [SI]) line maps observed with Herschel/PACS and
Spitzer/IRS is consistent with the excitation of these species in outflows. The
only exception is the emission from [CII], which is likely to be excited by UV
radiation from the nearby protostellar source SMM6.}

\item{Line emission in SMM3 is extended, following an SE-NW outflow pattern. For
SMM4, line emission shows one significant peak at $\sim 10\arcsec$ to the N-NW
from the protostar.}

\item{CO and H$_2$ trace gas at temperatures of $\sim$300 and $\sim$1000 K.
Within the available resolution, emission from both molecules is confined to the
same areas along outflows, and their emission pattern has very similar
morphological characteristics. These findings suggest that CO and H$_2$ are
excited by a common underlying mechanism.}

\item{H$_2$O and OH trace gas at rotational temperatures in the range
$\sim$100-200 K.  Even though these species are chemically related, the large
scatter of data points in the excitation analysis does not allow  us to tightly
constrain their properties.}

%[NJE: Can you claim first time given all the other references you gave?]
\item{The association between CO and H$_2$ allows us
 to directly
estimate the CO abundance in excited gas. The CO/H$_2$ ratio is found to vary
from 10$^{-5}$ at $\sim$300 K up to 2 $\times$ 10$^{-4}$ at $\sim$1000 K.}

\item{The existence of dissociative J-type shocks is strongly suggested by the
high atomic line fluxes as well as diagnostics of the [OI]/H$_2$ ratio. In
addition J-shocks can reproduce the observed H$_2$ emission and the CO emission
at $\sim$300 K. The lack of model predictions for J$_{up} \ge$20 
%does not allow to conclude on 
prevents a conclusive test of whether J-shocks can reproduce the 
warm gas traced by CO. All
J-shock diagnostics and model comparisons provide consistent pre-shock densities
of 2$\times10^{4}$ cm$^{-3}$, shock velocities from 15 to 25 km $^{-1}$ and beam
filling factors from 10\% to 90\%. This common set of shock parameters is
consistent with a common physical mechanism responsible for the excitation of CO
and H$_2$. J-shock models however, cannot sufficiently reproduce the observed
H$_2$O emission.}

\item{C-shocks can reproduce the observed emission for CO and H$_2$ only if two
temperatures and small beam filling factors are considered, suggesting a layered
temperature structure, possibly in a bow-shock. Non-dissociative shocks fit best
the H$_2$O emission. The predicted pre-shock densities and velocities vary
significantly for different species, in support of a projected 2-dimensional
layering.}

\item{The two-temperature structure for CO and H$_2$ has been observed in the
majority of outflows from protostellar sources. The current analysis indicates
that this structure is likely to reflect a physical break between C- and J-type
shocks.}

\end{itemize}

The physical association between CO and H$_2$ indicated here, needs to be also
confirmed in different protostellar environments. 
%A subset of the 
DIGIT embedded sources that have been mapped with Spitzer 
%is currently examined and 
will be analyzed in a follow-up paper. 
If this association is more general, 
then this larger sample will help to better constrain the ratio
between CO and H$_2$ in the warm gas around young protostars. In addition, more
complete and detailed shock models are required in order to control the
influence of different types of shocks on the abundances of both molecules.

\begin{acknowledgements}

This research was supported by a grant from the Instrument Center for Danish Astrophysics (IDA) and a Lundbeck Foundation Group Leader Fellowship to JKJ. Research at Centre for Star and Planet Formation is funded by the Danish National Research Foundation and the University of CopenhagenÕs programme of excellence. Support for this work, part of the Herschel Open Time Key Project Program, was provided by NASA through an award issued by the Jet Propulsion Laboratory, California Institute of Technology. We would like to thank the anonymous referee for his/her comments that greatly improved the manuscript.

 \end{acknowledgements}

\bibliographystyle{aa} \bibliography{serpens_ads}

\appendix \section{Spitzer spectral line maps of Serpens SE} \label{app:a}

\begin{figure*} \centering
\resizebox{\hsize}{!}{\includegraphics{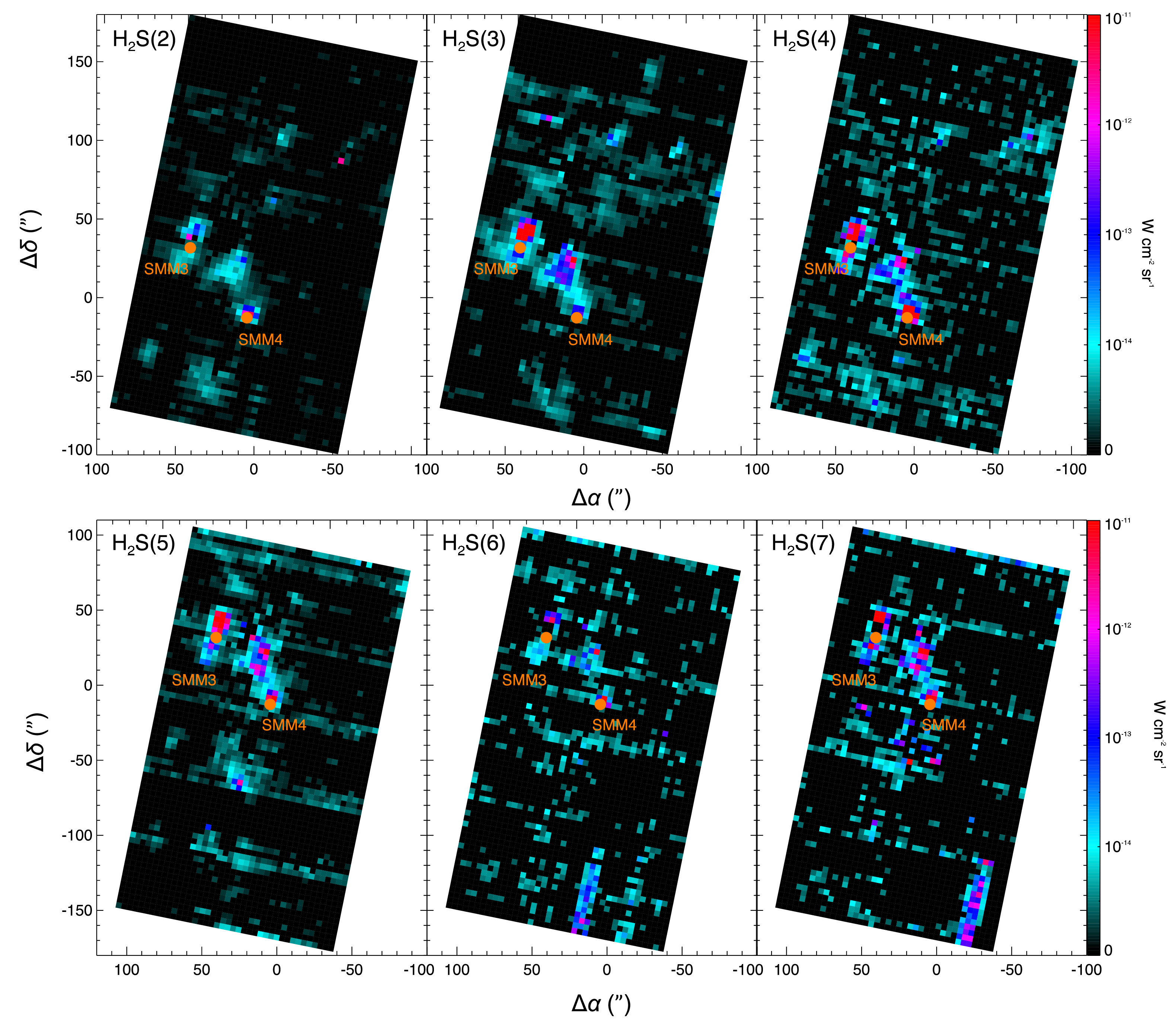}}
\caption{Spitzer/IRS H$_2$ maps obtained with the SL modules, at a resolution of
3.5$\arcsec$/spaxel} \label{fig:a1} \end{figure*}

\begin{figure*} \centering
\resizebox{\hsize}{!}{\includegraphics{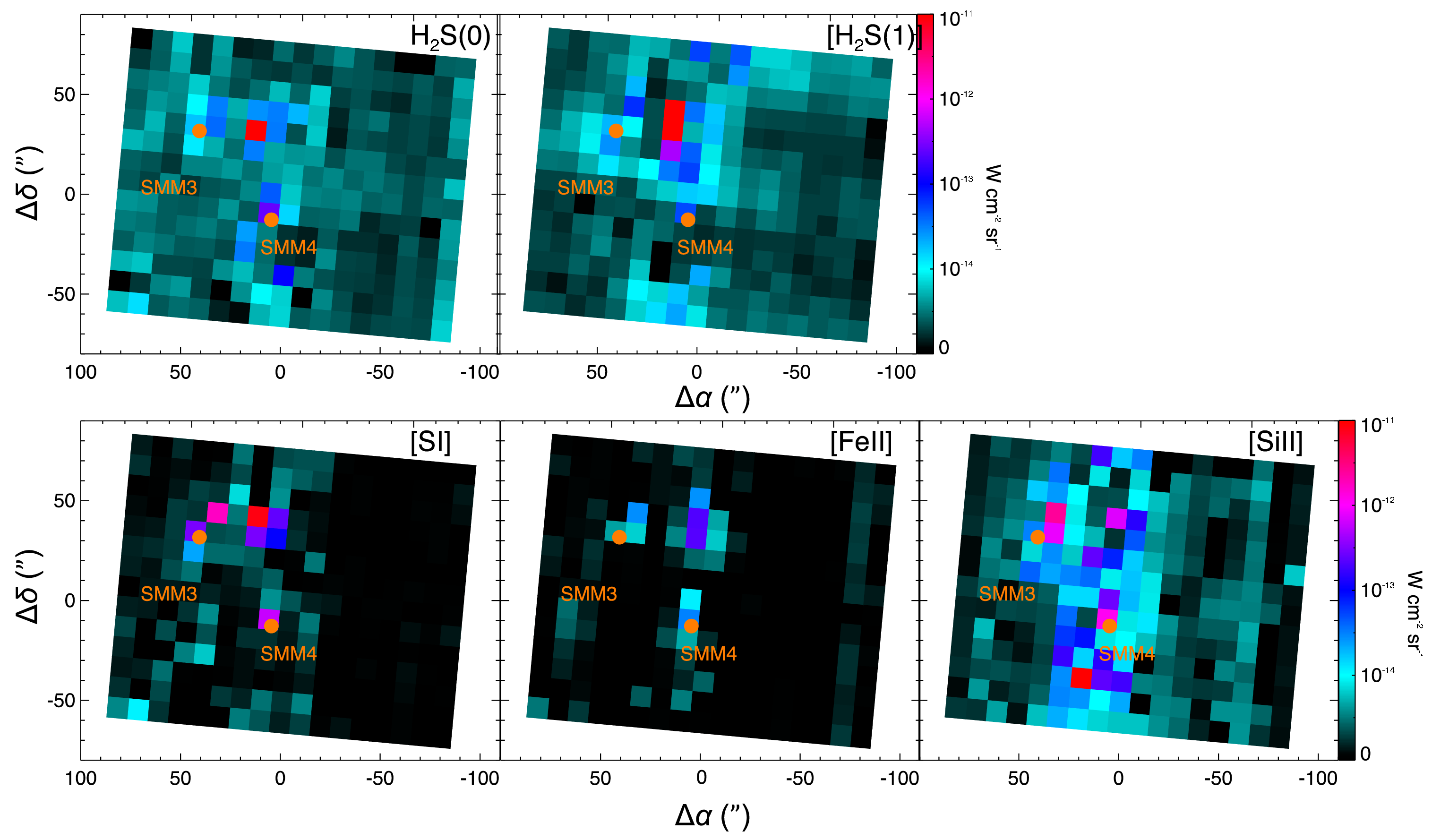}}
\caption{Spitzer/IRS line maps of the S(0)l and S(1) H$_2$ transitions, along
with atomic line maps from the LL modules at a resolution of
10.5$\arcsec$/spaxel} \label{fig:a2} \end{figure*}

Spitzer/IRS maps at their full extent and resolution. Maps of the pure
rotational H$_2$ lines S(2) - S(7) falling in the SL module wavelength range are
presented in figure \ref{fig:a1}. For reference, the positions of SMM3 and SMM4
are indicated with filled orange circles. In order to obtain maximum coverage,
both on and off positions were used for the compilation of the SL1 and SL2
module data-cubes.  This extended coverage appears to the N  or the S (SL2 and
SL1 modules, respectively) of SMM3 and SMM4. The resolution of the SL maps is
3.5$\arcsec$ per spaxel, and weak emission-like stripes are due to residual
rogue pixels during the reduction. Figure \ref{fig:a2} presents the S(1), S(0)
H$_2$ line maps, along with ones corresponding to atomic lines from [SI], [FeII]
and [SiII] (at 25.2, 26.0 and 34.8 $\micron$). The LL map resolution is
10.5$\arcsec$ per spaxel.

\section{Short description of the POMAC code} \label{app:b} The POMAC
(Poor-man's CLEAN) algorithm (Lindberg et al., in prep.) is a deconvolution
algorithm used to separate point sources from extended emission in PACS data.
The code is a modified version of the CLEAN algorithm (Hšgbom 1974), with the
difference that it requires the positions of testing "point" sources. Such
points in the present case are selected to be the spaxels displaying emission
maxima. The code produces instrument PSFs corresponding to point sources in each
pre-defined point source position for the wavelength of the spectral line (or
continuum emission) that is to be deconvolved. This is constructed by assuming
the Herschel telescope primary beam to be a Rayleigh-criterion Gaussian, which
is then overlaid on the PACS spaxel grid to measure how much of the Gaussian
that falls into each spaxel. 

After the PSFs have been determined, the CLEAN part of the code
commences. It studies the line (or continuum) emission in the PACS grid to find
the pre-defined point source responsible for the strongest flux in the field. A
fraction of this flux convolved with the PSF of that point source is subtracted
from the line map, and the subtracted flux is added to the cleaned flux of this
point source. This step is then repeated until certain stop criteria are met.
The remaining flux in the residual map is then composed of extended emission and
emission from any unknown point sources, and the cleaned line fluxes correspond
to that of the point sources. Note that the residual map is non-deconvolved, and
shall not be interpreted as a deconvolved map of the extended emission. Finally,
the sum of the cleaned flux and the residual flux is compared to the original
map as a consistency check. For a well-centered point source without extended
emission (such as the continuum emission of HD100546) the results of this method
agree well with the results from the PSF correction factor method (the
differences are less than 20\% across the PACS band).  

\section{Excitation diagrams} \label{app:c}

 \begin{figure*} \centering
\resizebox{\hsize}{!}{\includegraphics{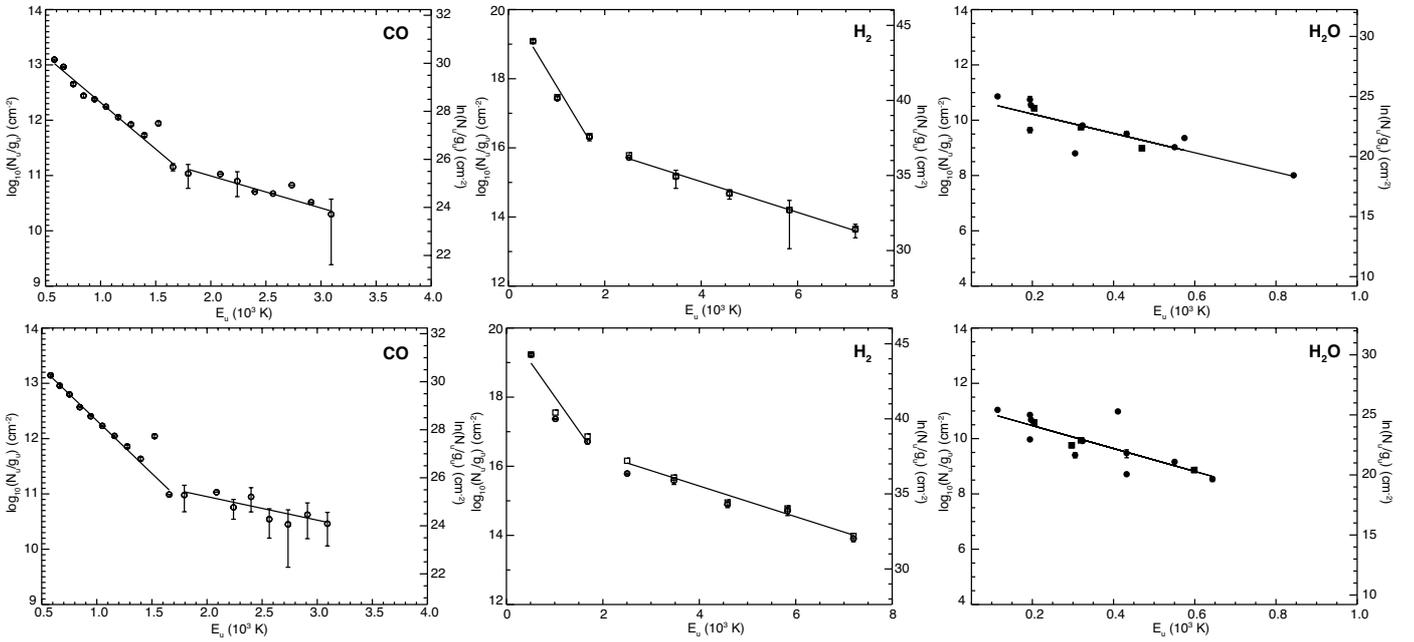}}
\caption{Excitation diagrams for CO, H$_2$ and H$_2$O at SMM3c and SMM3r (upper
and lower panels, respectively)} \label{fig:b1} \end{figure*}

 \begin{figure*} \centering
\resizebox{\hsize}{!}{\includegraphics{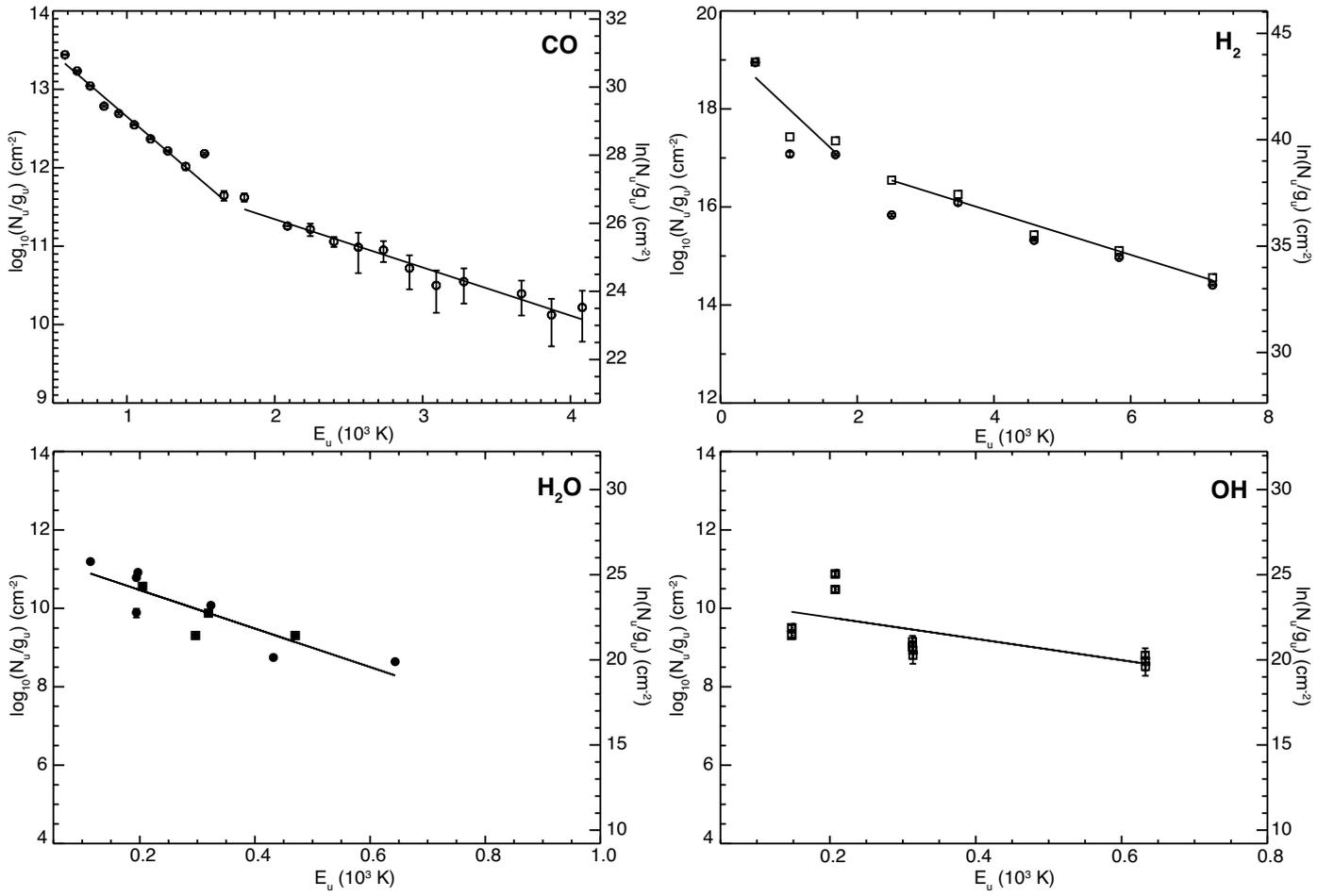}}
\caption{Excitation diagrams for CO, H$_2$, H$_2$O and OH at SMM4c}
\label{fig:b2} \end{figure*}

\end{document}